\documentclass[compsoc,conference,a4paper,10pt,times]{IEEEtran}
\IEEEoverridecommandlockouts
\usepackage{amsmath,amssymb,amsfonts}
\usepackage{graphicx}
\usepackage{textcomp}
\usepackage{bmpsize}
\usepackage[table]{xcolor}
\usepackage{lipsum}
\usepackage{afterpage}
\usepackage{pgfplots}
\usepackage{enumitem}

\def\BibTeX{{\rm B\kern-.05em{\sc i\kern-.025em b}\kern-.08em
    T\kern-.1667em\lower.7ex\hbox{E}\kern-.125emX}}
    
\usepackage{algorithm}
\usepackage{algpseudocode}
\usepackage{tabularx}
\usepackage[hyphens]{url}

\usepackage[colorlinks=true,urlcolor=black]{hyperref}
\usepackage{cite}
\algnewcommand{\LeftComment}[2][\algorithmicindent]{\State \hspace{#1} \(\triangleright\) #2}
\usepackage{tablefootnote}
\usepackage[bottom]{footmisc}

\pgfplotsset{compat=1.18} 

\begin{document}

\title{CovFUZZ: Coverage-based fuzzer for 4G\&5G protocols}

\author{\IEEEauthorblockN{1\textsuperscript{st} Ilja Siroš}
\IEEEauthorblockA{\textit{COSIC, KU Leuven} \\
Leuven, Belgium \\
ilja.siros@esat.kuleuven.be }
\and
\IEEEauthorblockN{2\textsuperscript{nd} Dave Singelée}
\IEEEauthorblockA{\textit{COSIC, KU Leuven} \\
Leuven, Belgium \\
dave.singelee@esat.kuleuven.be}
\and
\IEEEauthorblockN{3\textsuperscript{rd} Bart Preneel}
\IEEEauthorblockA{\textit{COSIC, KU Leuven} \\
Leuven, Belgium \\
bart.preneel@esat.kuleuven.be}
}
\maketitle

\begin{abstract}
4G and 5G represent the current cellular communication standards utilized daily by billions of users for various applications. Consequently, ensuring the security of 4G and 5G network implementations is critically important. This paper introduces an automated fuzzing framework designed to test the security of 4G and 5G attach procedure implementations. Our framework provides a comprehensive solution for uplink and downlink fuzzing in 4G, as well as downlink fuzzing in 5G, while supporting fuzzing on all layers except the physical layer. To guide the fuzzing process, we introduce a novel algorithm that assigns probabilities to packet fields and adjusts these probabilities based on coverage information from the device-under-test (DUT). For cases where coverage information from the DUT is unavailable, we propose a novel methodology to estimate it. When evaluating our framework, we first run the random fuzzing experiments, where the mutation probabilities are fixed throughout the fuzzing, and give an insight into how those probabilities should be chosen to optimize the \textit{Random fuzzer} to achieve the best coverage. Next, we evaluate the efficiency of the proposed coverage-based algorithms by fuzzing open-source 4G stack (srsRAN) instances and show that the fuzzer guided by our algorithm outperforms the optimized \textit{Random fuzzer} in terms of DUT's code coverage. In addition, we run fuzzing tests on 12 commercial off-the-shelf (COTS) devices. In total, we discovered vulnerabilities in 10 COTS devices and all of the srsRAN 4G instances.
\end{abstract}

\begin{IEEEkeywords}
Fuzzing, 4G, LTE, 5G, NR, srsRAN, Open5Gs
\end{IEEEkeywords}

\section{Introduction}
The 4G and 5G standards are widely adopted wireless communication protocols that provide high-speed data transmission to mobile devices. According to recent estimates~\cite{GSMA}, there are over $5$ billion unique mobile subscribers and more than $12$ billion mobile connections in 2024. However, it is important to recognize that the 3rd Generation Partnership Project (3GPP)~\cite{3GPP} specifications, which define the 4G and 5G standards, are extensive and complex. This complexity can lead to inadvertent errors during implementation, potentially introducing vulnerabilities into the system.

This paper introduces a  fuzzing framework designed to identify vulnerabilities in 4G and 5G network implementations that can arise during the \textit{Attach procedure}. It is versatile, accommodating 4G uplink and downlink as well as 5G downlink communication scenarios, effectively simulating the behavior of malicious 4G user equipment or a 4G/5G base station, respectively. In our framework we aim to avoid the excessive human effort in the test case generation process, instead relying on automated fuzzing techniques. Our framework does not aim to uncover logical bugs, such as instances where the \textit{DUT}'s behavior deviates from specifications. Instead, it focuses on identifying bugs that cause crashes or hangs of the \textit{DUT}.

Most modern basebands are implemented in C/C++, which is inherently not memory safe. This can lead to memory-safety vulnerabilities, such as \textit{use-after-free}~\cite{CWE-416} and various \textit{buffer overflow}~\cite{CWE-120},~\cite{CWE-121},~\cite{CWE-122} vulnerabilities. These vulnerabilities are critical and can potentially be exploited to achieve remote code execution on the baseband~\cite{CVE-2022-21744,CVE-2023-21517}.  Therefore, the primary objective of our fuzzing framework is to automatically uncover memory-safety vulnerabilities. However, it can also detect other types of bugs, such as \textit{reachable assertion}s~\cite{CWE-617}, which can lead to denial-of-service attacks~\cite{DoS}. While these examples are not exhaustive, they illustrate the types of vulnerabilities our framework aims to uncover.

Our fuzzing framework is built upon open-source 4G and 5G implementations. For 4G fuzzing, we use srsRAN 4G~\cite{SRSRAN_4G}, while for 5G fuzzing, we utilize the srsRAN Project~\cite{srsRAN_Project} along with the 5G Core Network from Open5Gs~\cite{Open5Gs}. When fuzzing (network) protocols, generating valid inputs is one of the most challenging tasks. To address this, we leverage on the open-source stack instances to generate benign packets, which we then mutate. These benign packets can be intercepted at two layers, namely RRC and MAC, allowing for real-time packet manipulation. Additionally, having multiple interception points mitigates the challenge of mutating integrity-protected or encrypted fields, as manipulations are performed on a higher layer than where integrity protection and encryption occur.

To mutate the packets, we introduce the concept of mutation probabilities assigned to every field in every packet. Based on this concept, we implemented \textit{Random} and \textit{Coverage-based fuzzers}. The \textit{Random fuzzer} serves as a baseline in our evaluation, operating without any guidance. However, we conducted experiments to determine the optimal initial field mutation probabilities, as selecting incorrect values can result in inefficient fuzzing. Throughout this paper, we refer to the \textit{Random fuzzer} with optimally chosen initial probabilities as the optimized \textit{Random fuzzer}.

The main contribution of our paper is the \textit{Coverage-based fuzzer}, which enhances the fuzzing process by dynamically adjusting field mutation probabilities based on the coverage feedback from the \textit{DUT}. We provide a detailed description of the novel algorithm behind this probability adjustment and thoroughly analyze how the algorithm's hyperparameters influence the fuzzer's performance. We consider both grey- and black-box scenarios, where the coverage information from the \textit{DUT} is available and not available, respectively. In the latter case, we propose a way to estimate the \textit{DUT}'s coverage information based on the coverage information from the open-source implementation of the base station and user equipment used for the benign packet generation for the downlink and uplink fuzzing, respectively.

We evaluate our novel algorithm using downlink and uplink fuzzing of srsRAN 4G instances, relying on coverage metrics to compare the efficiency of the proposed fuzzers. As a proof of concept, we also conducted fuzzing tests on twelve COTS 4G and 5G devices. We discovered vulnerabilities in ten of these devices, and in all srsRAN 4G instances.

In summary, our paper makes the following contributions: 
\begin{itemize}[leftmargin=*] 
    \item We present a fuzzing framework based on the srsRAN and Open5Gs open-source projects for 4G uplink and downlink fuzzing, as well as 5G downlink fuzzing of the \textit{Attach procedure}. Our framework can mutate packets across all 4G and 5G protocol stack layers except the physical layer.
    \item We introduce the concept of field mutation probabilities and implement two fuzzers: \textit{Random fuzzer} and \textit{Coverage-based fuzzer}. We conduct extensive experiments in both downlink and uplink scenarios to identify the optimal initial probabilities for the \textit{Random fuzzer}, which serves as a baseline for evaluating the performance of our coverage-based algorithm.
    \item For the \textit{Coverage-based fuzzer}, we propose a novel grey-box fuzzing algorithm that efficiently utilizes coverage feedback to fuzz 4G and 5G networks. We demonstrate that this algorithm can also be successfully executed in a black-box scenario without getting coverage information from the \textit{DUT}, but instead estimating it from the code coverage of an open-source instance responsible for generating benign packets.
    \item We evaluate the efficiency of the fuzzers implemented in our framework using the code coverage metric. Our results show that the \textit{Coverage-based fuzzer} improves the code coverage of the \textit{DUT} in both grey-box (by $47.6\%$ for downlink and $11.9\%$ for uplink fuzzing) and black-box (by $23.9\%$ for downlink and $11.3\%$ for uplink fuzzing) scenarios compared to the optimized \textit{Random fuzzer}.
    \item During our tests, we discovered vulnerabilities in 10 COTS devices and in all srsRAN 4G instances.
\end{itemize}

\noindent\textbf{Responsible Disclosure and Open Sources}. 
As part of responsible disclosure, we have notified the relevant UE manufacturers regarding our findings. To facilitate further research and investigations in this field, we have made our code available at~\url{https://anonymous.4open.science/r/CovFUZZ_4G_5G-9258}.

\section{4G\&5G background}
4G and 5G are the fourth- and fifth-generation cellular network technology standards developed by 3GPP. The architecture of 4G/5G networks can be broadly categorized into three main components: the \textit{User Equipment} (UE), the \textit{Radio Access Network} (RAN) comprising eNodeBs (ENB) in 4G and gNodeBs (GNB) in 5G, and the \textit{Core Network} (CN) which includes the \textit{Evolved Packet Core} (EPC) in 4G and the \textit{5G Core Network} (5GC) in 5G.

The UE represents an end device that utilizes 4G/5G for communication purposes. It connects to the RAN, which then links to the CN. The Core Network handles subscriber authentication and Internet access. Transmissions from the UE to the RAN/CN are called ``uplink" (UL), and from the RAN/CN to the UE -- ``downlink" (DL).

4G and 5G networks consist of two main components: the control plane, which manages operations like connection and mobility management, and the user plane, which handles user data traffic. As our framework aims to fuzz the \textit{Attach procedure}, it focuses solely on the control plane of both networks.

\subsection{4G/5G control plane stack}
\begin{figure}
    \centering
    \includegraphics[width=\columnwidth]{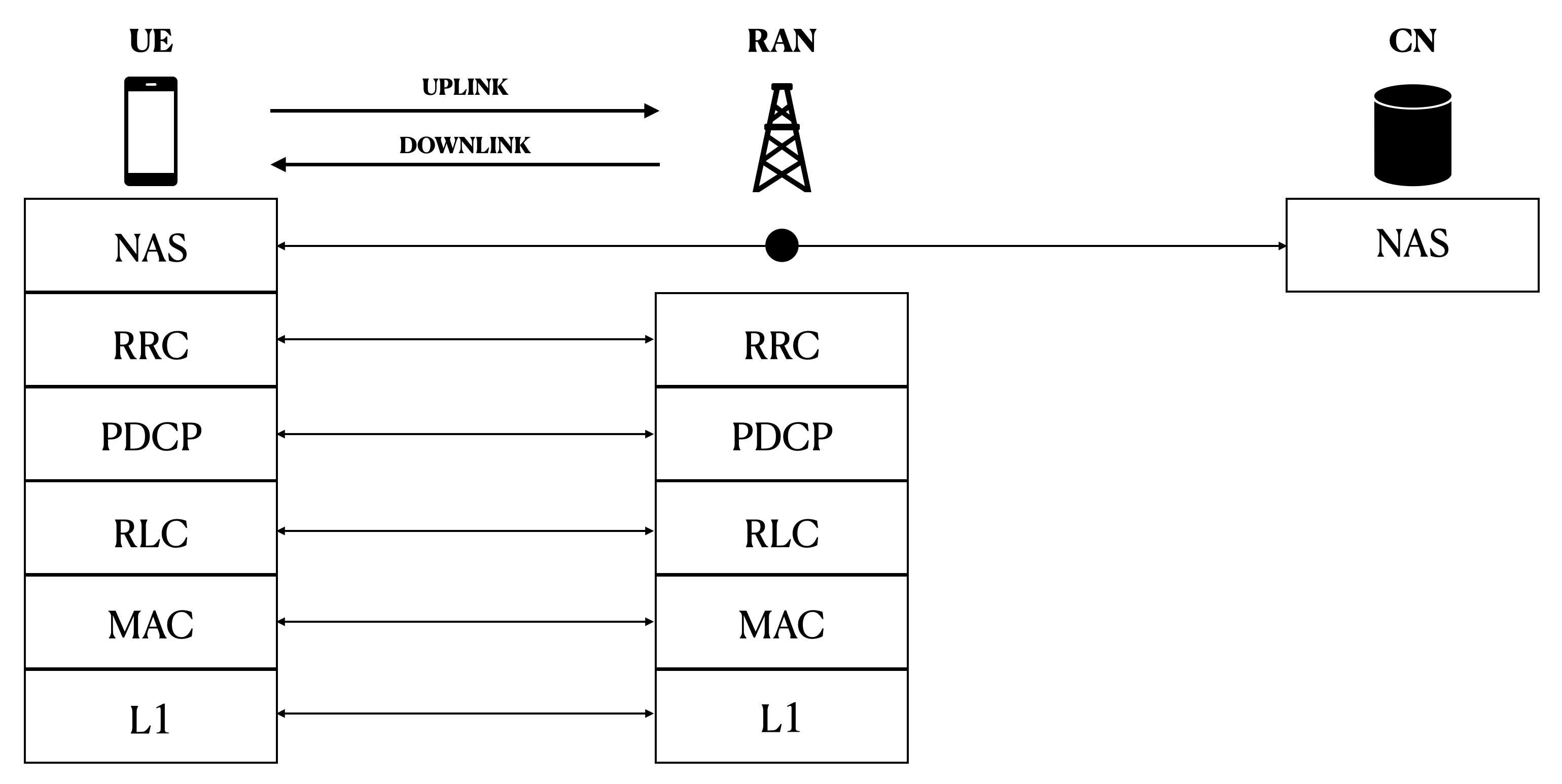}
    \caption{4G/5G control plane stack.}
    \label{fig:protocol_stack}
\end{figure}

The control plane protocol stack, depicted in Figure~\ref{fig:protocol_stack}, consists of several layers that collectively enable the functionality of the 4G/5G system.

Layer $1$, the Physical layer, handles the transmission of information across the air interface.

Layer $2$ includes the Medium Access Control (MAC), Radio Link Control (RLC), and Packet Data Convergence Control (PDCP) layers. The MAC schedules data, the RLC ensures error correction and packet segmentation, and the PDCP manages tasks like header compression and encryption

Layer $3$ consists of the Radio Resource Control (RRC) and Non-Access Stratum (NAS) layers. The RRC manages the connection between the user equipment (UE) and the radio access network (RAN), while the NAS oversees communication between the UE and the core network, handling session management and ensuring continuous connectivity.

\subsection{4G/5G channels}

In the 4G/5G protocol stack, information is transmitted through logical, transport, and physical channels. Logical channels define the type of data transmitted, such as the Broadcast Control Channel (BCCH) for system-wide broadcasting, the Paging Control Channel (PCCH) for paging information, the Common Control Channel (CCCH) for random access, and the Dedicated Control Channel (DCCH) for user-specific control data. Transport channels manage data transfer between the MAC and physical layers, while physical channels handle the actual transmission. In our framework, we focus on fuzzing packets transmitted via the CCCH and DCCH channels, which cover all packets in the \textit{Attach procedure}.

\subsection{Attach procedure}
\begin{figure}
    \centering
    \includegraphics[width=\columnwidth]{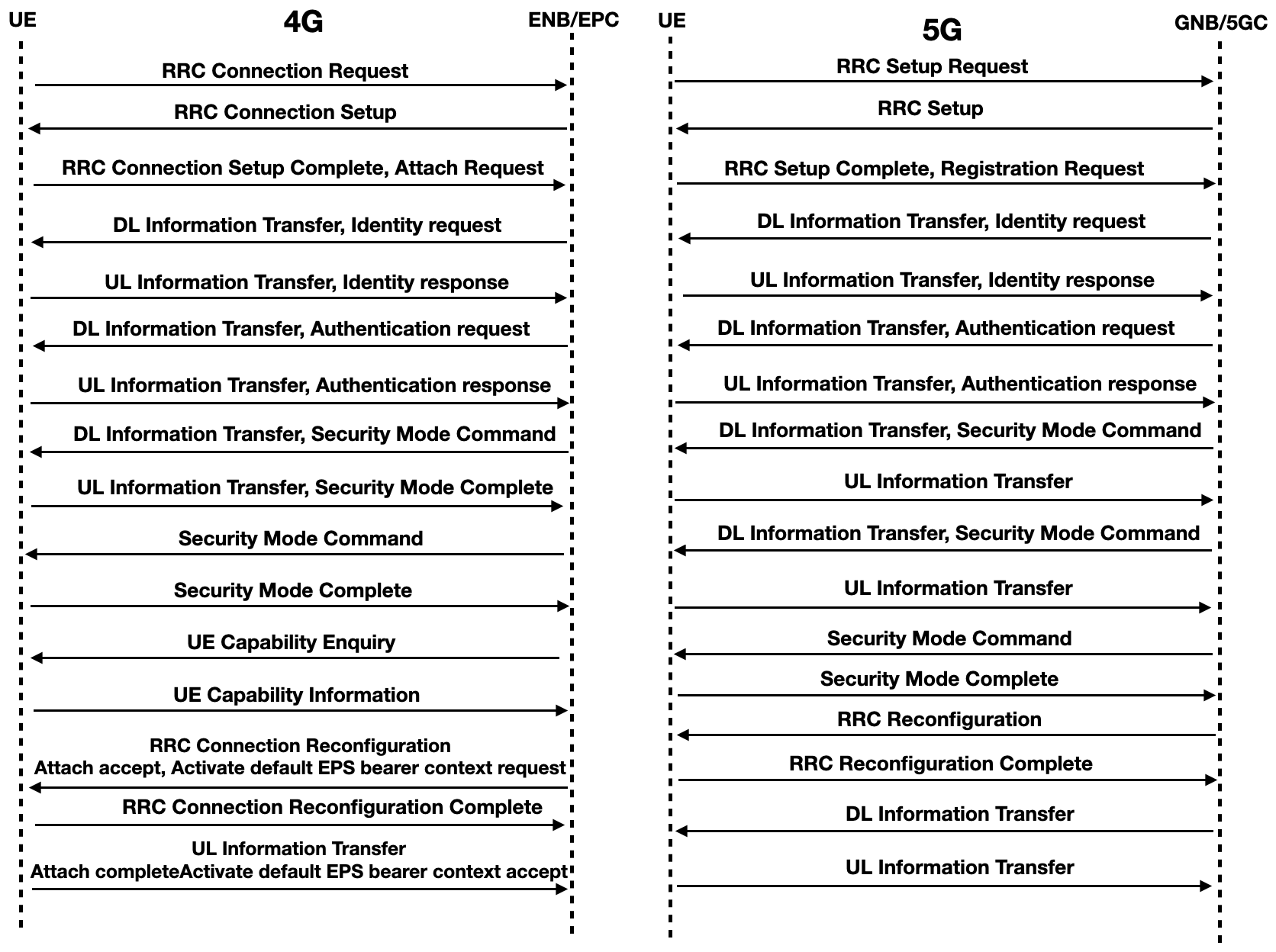}
    \caption{Attach procedure in 4G and 5G.}
    \label{fig:attach_procedure}
\end{figure}

Our research focuses on fuzzing the attach procedure. In brief, the attach procedure involves mutual authentication, UE registration with the network, and the establishment of a bearer between the UE and the CN.
Figure~\ref{fig:attach_procedure} demonstrates a packet flow during a (successful) \textit{Attach procedure}. It is important to note that other messages may also occur during the attach procedure, such as when the attach is rejected or the radio link connection is broken. Even when encountering these messages, our approach remains consistent in fuzzing packets that flow through logical channels CCCH and DCCH.

\section {Fuzzing framework architecture}
Our fuzzing framework supports both downlink and uplink 4G, as well as downlink 5G fuzzing. The limitation for 5G uplink fuzzing arises from the fact that, at the time of this research, the 5G version of the srsUE in the srsRAN project lacked the turn off/on functionality needed to reset and renew fuzzing iterations.
 
\begin{figure*}
    \centering
    \includegraphics[width=\textwidth]{"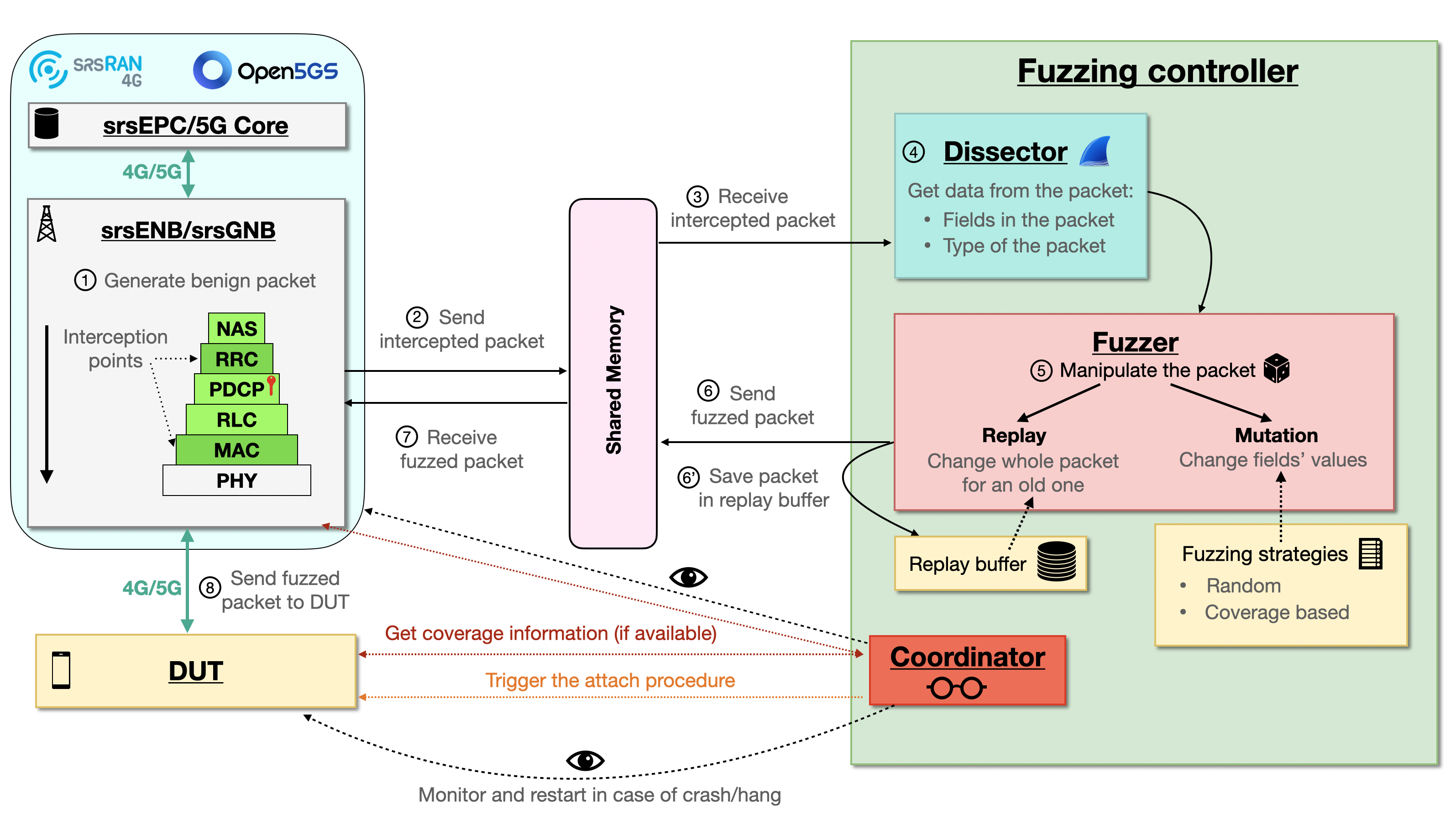"}
    \caption{The illustration of our fuzzing framework architecture for the downlink 4G/5G fuzzing. The steps shown in the figure outline the fuzzing flow for one packet. The architecture for uplink 4G fuzzing is similar and it can be achieved by inverting the left part of the figure, i.e. the interception points are in srsUE, and srsENB/srsEPC serve as the \textit{DUT}(s).}
    \label{fig:framework_architecture}
\end{figure*}

The fuzzing framework consists of three primary components: the \textit{4G/5G protocol stack implementation}, \textit{DUT}, and \textit{Fuzzing controller}. In this section, we provide a detailed description of each component.
Figure~\ref{fig:framework_architecture}
 illustrates the architecture of the downlink 4G/5G fuzzing framework.

\subsection{4G/5G Protocol stack implementation}\label{protocol_stack_impl}
In our fuzzing framework, the 4G/5G protocol stack implementation is responsible for generating benign packets. We utilize the srsRAN 4G suite~\cite{SRSRAN_4G} for generating benign 4G packets and the combination of srsRAN Project~\cite{srsRAN_Project} with Open5GS~\cite{Open5Gs} for 5G. Specifically, for 4G, we use srsUE for uplink as well as srsENB and srsEPC for downlink packet generation. Similarly, for 5G, we employ srsGNB and Open5GS CN to generate benign downlink packets.
We have also integrated the following additional functionalities into these open-source instances:

\subsubsection{Packet interception hooks}
We implemented packet interception hooks at two layers, RRC and MAC, by integrating custom code into the srsGNB, srsENB, and srsUE implementations for downlink 5G, downlink 4G, and uplink 4G interception, respectively. This custom code is a hook that sends intercepted packets to the \textit{Fuzzing controller} via a low-latency shared memory interface and suspends the protocol flow until the fuzzed message is processed and returned. Since encryption and integrity protection occur at the PDCP layer, our interception allows manipulation of NAS and RRC fields at the RRC layer and PDCP, RLC, and MAC fields at the MAC layer.

\subsubsection{LLVM address and coverage instrumentation}
We instrumented the srsRAN instances (i) with the LLVM address sanitizer~\cite{LLVM_Address_Sanitizer} to ensure the detection of memory errors and (ii) with the LLVM coverage sanitizer~\cite{LLVM_Coverage_Sanitizer} to obtain an insight into the code coverage of the srsRAN instances. The purpose of gathering coverage information is twofold. First, we use it as a metric to evaluate the performance of the fuzzers in our framework. Second, we use it to guide the fuzzing process, aiming to maximize the coverage of the \textit{DUT}.

\subsubsection{Changes in srsUE}\label{srsUE_Changes}
We added a custom code to srsUE to speed up the resetting process of the UE between fuzzing iterations. This code includes a TCP socket listener that can be activated by the \textit{Fuzzing controller} as needed. For the TCP socket implementation, we utilized the ZMQ library~\cite{ZeroMQ}.
Additionally, to enhance the efficiency of the fuzzing process, we decrease the value of the timer T$3410$, which defines the maximum duration between \textit{Attach request} and \textit{Attach accept} messages, from default $15$ to $1.5$ seconds. Finally, we decreased the number of maximum attach attempts to $1$. This modification ensures that if the attach procedure fails once, srsUE does not automatically attempt to attach to the 4G network again. Therefore, each fuzzing iteration precisely aligns with one attach procedure.

\subsubsection{Changes in srsEPC and Open5Gs} \label{srsEPC_Open5Gs}
After the UE successfully connects to a 4G or 5G network, subsequent connection attempts involve a lighter attach procedure without NAS signaling messages, as the CN already recognizes the UE. This negatively impacts fuzzing, as it limits the variety of messages available for testing. To address this issue, we implemented patches for srsEPC and Open5Gs to ensure that the CN ``forgets'' about the UE after an attach procedure is completed. This way, NAS messages are present during every fuzzing iteration, enhancing the effectiveness of the fuzzing process.

In addition, we developed a patch for Open5Gs to enable support for the XOR authentication algorithm, which is commonly used by many test SIM cards, including ours. This patch ensures compatibility with these SIM cards during the fuzzing experiments.

\subsection{DUT}
The \textit{DUT} is our fuzzing target. In the case of the downlink fuzzing for both 4G and 5G, the \textit{DUT} can be various devices such as mobile phones, IoT devices, simulated UEs like srsUE, or standalone 4G/5G modems. For uplink fuzzing in 4G, the \textit{DUT} comprises the ENB or EPC.\footnote{Note that one can easily use the principles explained in this paper to build an uplink 5G fuzzer and fuzz the GNB/5GC}

This paper first evaluates the fuzzing framework on the open-source srsRAN 4G instances. In this setting, for the downlink fuzzing, srsUE serves as the \textit{DUT}; for the uplink fuzzing, srsENB and srsEPC function as the \textit{DUT}s. By conducting evaluations with these open-source \textit{DUT}s, we can thoroughly test our grey-box and black-box coverage-based fuzzing algorithms. Once these evaluations are completed, we apply the fuzzing framework to COTS 4G/5G UEs.

\subsection{Fuzzing controller}
\textit{Fuzzing controller} is a critical linking component that effectively coordinates the entire fuzzing process. Among its primary responsibilities are the following:
\begin{itemize}
    \item Receive the intercepted packets via the shared memory interface and send the fuzzed packets back.
    \item Log the fuzzing process.
    \item Monitor the status of all framework components and restart them if needed.
    \item Initiate and terminate the fuzzing process.
    \item Gather the coverage feedback from the 4G/5G protocol stack components if available.
\end{itemize}
The whole fuzzing flow in \textit{Fuzzing controller}'s perspective can be seen in Algorithm~\ref{alg:fuzzing_flow}.

\begin{algorithm}
    \caption{Fuzzing flow}\label{alg:fuzzing_flow}
    \begin{algorithmic}[1]
        \State \textbf{Input:} $max\_i$ \Comment{Maximum number of iterations}
        \State
        \For {$i$ in $1,2,\dots, max\_i$}
            \LeftComment[0\dimexpr\algorithmicindent]{Send attach to network command to UE}
            \State \Call{TriggerAttachProcedure}{$ $}
            \LeftComment[0\dimexpr\algorithmicindent]{While attach procedure is not finished}
            \While {\texttt{NotAttachOver}}
            \LeftComment[0\dimexpr\algorithmicindent]{Receive the packet via shared memory}
                \State $packet \gets \ $\Call{SHM.recv}{$ $} %\Comment{Receive the packet via shared memory}
            \LeftComment[0\dimexpr\algorithmicindent]{Dissect the packet}
                \State \Call{dissect}{$packet$}
            \LeftComment[0\dimexpr\algorithmicindent]{Fuzz the packet}
                \State \Call{fuzzer.fuzz}{$packet$}
            \LeftComment[0\dimexpr\algorithmicindent]{Send the packet back to interception point}
                \State \Call{SHM.send}{$packet$}
            \EndWhile
            \LeftComment[0\dimexpr\algorithmicindent]{Check if any instances crashed}
            \If {not AllComponentsAlive}
                \LeftComment[0\dimexpr\algorithmicindent]{Save the seed that caused a crash}
                \State \Call {SaveCurrentSeed}{$ $}
                \LeftComment[0\dimexpr\algorithmicindent]{Handle the crash}
                \State \Call {HandleCrash}{$ $}
            \EndIf
            \LeftComment[0\dimexpr\algorithmicindent]{Receive information about the code coverage}
            \State \Call {GetCoverage}{$ $}
            \LeftComment[0\dimexpr\algorithmicindent]{Let fuzzer know that iteration has finished}
            \State \Call {fuzzer.PrepareNextIter}{$ $}
            \If {fuzzer.NeedToFinish}
                \LeftComment[0\dimexpr\algorithmicindent]{Finish as the fuzzing strategy is completed}
                \State \textbf{break}
            \EndIf
            \LeftComment[0\dimexpr\algorithmicindent]{Disconnect UE from the network and reset stack}
            \State \Call {ResetUEStack}{$ $}
        \EndFor
        
    \end{algorithmic}
\end{algorithm}

The main sub-components of \textit{Fuzzing controller} are \textit{Dissector}, \textit{Fuzzer} and \textit{Coordinator}.

\subsubsection{Dissector}
The \textit{Dissector} is responsible for packet dissection, which involves identifying the packet type and extracting the fields it contains. We utilize a custom library developed on top of the Wireshark~\cite{Wireshark} libraries for this purpose. The same library was used by Shang et al.~\cite{U-Fuzz}.

\subsubsection{Fuzzer}
The \textit{Fuzzer} is an abstract class responsible for mutating packets according to its specific strategy. Within our framework, we have implemented two fuzzers detailed in the following section. Furthermore, we designed our \textit{Fuzzer} to be highly generic, enabling its application to other wireless protocols with ease.
Each fuzzer operates on \textit{Packets}, which consist of various \textit{Fields}.

The \textit{Packet} is a class designed to encapsulate an intercepted 4G/5G packet within our framework. It includes the following attributes: raw packet bytes, intercepted layer information, transmission direction (UL or DL), and some additional details provided by the dissector, such as the packet type and an array of \textit{Field}s contained within it.

The \textit{Field} class represents an individual field within a \textit{Packet}. It contains the following attributes: a name of the field, an index of the field that is used to distinguish fields with the same name, and a location of the field represented with the following three variables:
\begin{itemize}
    \item \textbf{Offset} indicates the offset of the field's leftmost byte in the packet.
    \item \textbf{Length} indicates the number of bytes the field spans over.
    \item \textbf{Mask} indicates which bits contribute to the resulting field's value. This is necessary because fields in a packet are often not byte-aligned.
\end{itemize}

\begin{figure}
    \centering
    \includegraphics[width=\columnwidth]{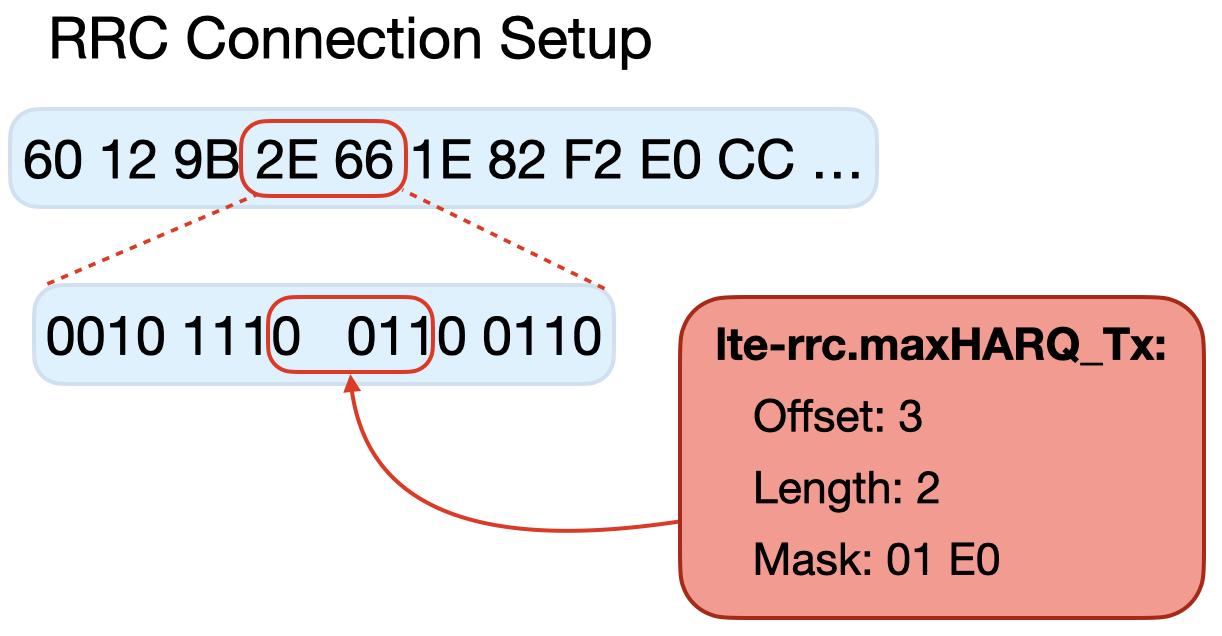}
    \caption{An example of extracting the value of the \textit{lte-rrc.maxHARQ\_Tx} field from the intercepted 4G RRC Connection Setup packet.}
    \label{fig:extract_field_value}
\end{figure}

An example of how those three variables uniquely identify the field in the packet is shown in Figure~\ref{fig:extract_field_value}.

For the purpose of fuzzing, we assign a mutation probability to each \textit{Field} within every packet. Details on how these probabilities are utilized during the fuzzing process are discussed in Section~\ref{sec:fuzzers}.

\subsubsection{Coordinator}
The \textit{Coordinator} component plays a pivotal role in managing the fuzzing process and handling information from both the \textit{DUT} and the \textit{protocol stack implementation}. Its main tasks can be described as follows.

\textbf{Monitoring the \textit{DUT} and the\textit{ protocol stack implementation}.} The \textit{Coordinator} is responsible for monitoring both the \textit{DUT} and the \textit{protocol stack implementation} to detect any instances of misbehavior such as crashes or hangs. A crash in a software instance (srsRAN or Open5Gs) is detected by checking if the corresponding process is still running. For the COTS UE, however, we rely mainly on the AT commands to check its aliveness. 

\textbf{Triggering start/end of fuzzing iterations.} In our framework, initiating a fuzzing iteration is synonymous with triggering the attach procedure. The process for starting and ending these iterations varies depending on whether we are working with srsUE or a COTS UE. When using srsUE, the \textit{Coordinator} initiates the attach procedure by leveraging TCP sockets, as detailed in Section~\ref{srsUE_Changes}.
For COTS UEs, the \textit{Coordinator} controls the initiation and termination of the attach procedure by toggling airplane mode. That can be accomplished using the appropriate AT command, i.e. \textit{AT+CFUN=4} to activate airplane mode and \textit{AT+CFUN=1} to deactivate it. Alternatively, for Android smartphones that do not expose an access to the AT commands, this can be achieved using the Android Debug Bridge (ADB)~\cite{adb}, by executing the \textit{adb shell cmd connectivity airplane-mode enable/disable} commands. By relying exclusively on airplane mode functionality, we ensure that the vast majority of the COTS UEs can be fuzzed using our framework.

\textbf{Gather the coverage information from the \textit{DUT} (if available) and \textit{protocol stack implementation} instances.} In our framework we rely on the code coverage to guide our fuzzer. Following each fuzzing iteration, the \textit{Coordinator} collects information about the code coverage from instances where this data is accessible. Details on how this coverage information is used to refine and optimize our fuzzing strategies are elaborated in Section~\ref{Coverage-based_fuzzer}.

\section{Fuzzing: main concepts} \label{sec:fuzz_intro}
This section introduces the fundamental concepts utilized in our fuzzing framework.
Based on the definitions introduced in the preceding section, throughout this section, we employ the following notation:
\begin{itemize}
    \item $P$ - \textit{Packet}
    \item $f$ - \textit{Field}
    \item $F_P$ - set of \textit{Field}s in the \textit{Packet} $P$
    \item $V_f$ - set of values the \textit{Field} $f$ can take
    \item $p^i_f$ - probability of the \textit{Field} $f$ to be mutated after $i$-th fuzzing iteration
\end{itemize}

Our framework supports two methods of manipulating the \textit{Packet}: \textit{Replay} and \textit{Mutation}. \textit{Replay} involves replacing the entire \textit{Packet}, whereas \textit{Mutation} targets individual \textit{Fields} within the \textit{Packet}. To mutate multiple \textit{Fields} within the \textit{Packet}, multiple \textit{Mutations} are applied.
We encapsulate \textit{Replay} and \textit{Mutation} functionalities within the \textit{Patch} class, which manipulates individual \textit{Packets}. A \textit{Seed} is then formed from the set of \textit{Patches} applied to the packets in each fuzzing iteration. Below, we provide a detailed description of each method.

\subsection{Manipulating the packet}
\subsubsection{Replay}
The concept of \textit{Replay} involves fully replacing an intercepted \textit{Packet} by replaying one of the previously captured \textit{Packet}s. One limitation in our framework is that both \textit{Packets}—the intercepted one and the replayed one—must share the same logical channel. To facilitate \textit{Replay}, we maintain a replay buffer, organized as a map with channel and layer as keys, and a set of corresponding \textit{Packets} as values.
When applying a \textit{Replay} to an intercepted \textit{Packet}, we select a suitable \textit{Packet} from the replay buffer. Rather than choosing randomly, we prioritize more recent entries in the buffer. This approach enhances the likelihood that the replayed \textit{Packet} aligns with any encryption or integrity protection expected by the \textit{DUT}.

\subsubsection{Mutation}
The \textit{Mutation} class is responsible for mutating one \textit{Field} in the \textit{Packet}. \textit{Mutation} encompasses the \textit{Field} to mutate and a \textit{Mutator} to be applied.

\textit{\textbf{Mutator}} is a simple strategy aimed to mutate the value of a \textit{Field}. The implemented \textit{Mutators} in our framework include:
\begin{itemize}
    \item RAND - sets value of a \textit{Field} to a random value
    \item MAX - sets value of a \textit{Field} to maximum
    \item MIN - sets value of a \textit{Field} to minimum
    \item ADD - adds one to the current value
    \item SUB - subtracts one from the current value
    \item SET X - sets the value of a \textit{Field} to predefined value X
\end{itemize}
When selecting a \textit{Mutator}, our framework randomly chooses from RAND, MAX, MIN, ADD and SUB. Once a \textit{Mutator} is selected, the corresponding \textit{Field} within the \textit{Packet} is identified, its current value is adjusted according to the chosen \textit{Mutator}, and the updated value is then applied back to the \textit{Packet}. After a successful mutation, the \textit{Mutator} type is modified to SET X, where X is the new value.
Suppose we want to repeat the same mutation another time. In that case, we apply the SET X \textit{Mutator} instead of the original one, ensuring that we end up with the same value for the mutated \textit{Field} as the last time.

\subsection{Creating a Patch}
The \textit{Patch} class serves as a fundamental component for facilitating packet fuzzing within our framework. It operates at the packet level and supports two distinct types of patches:
\begin{itemize}
    \item \textbf{\textit{MutationPatch}} contains a set of \textit{Mutations} to be applied to the packet. When applying a \textit{MutationPatch}, each mutation within the patch is iteratively applied to the packet.
    
    \item \textbf{\textit{ReplayPatch}} contains an instance of the \textit{Replay} class. When applied to an intercepted \textit{Packet}, the \textit{ReplayPatch} simply replaces the intercepted packet with a previously captured packet using the \textit{Replay} mechanism.
    
\end{itemize}
The \textit{Patch} abstraction provides us an easy way to combine \textit{Replays} and \textit{Mutations} for fuzzing the packet by calling the \textit{Patch}'s \textit{apply} method.

\subsection{Creating a Seed}
\begin{figure}
    \centering
    \includegraphics[width=\columnwidth]{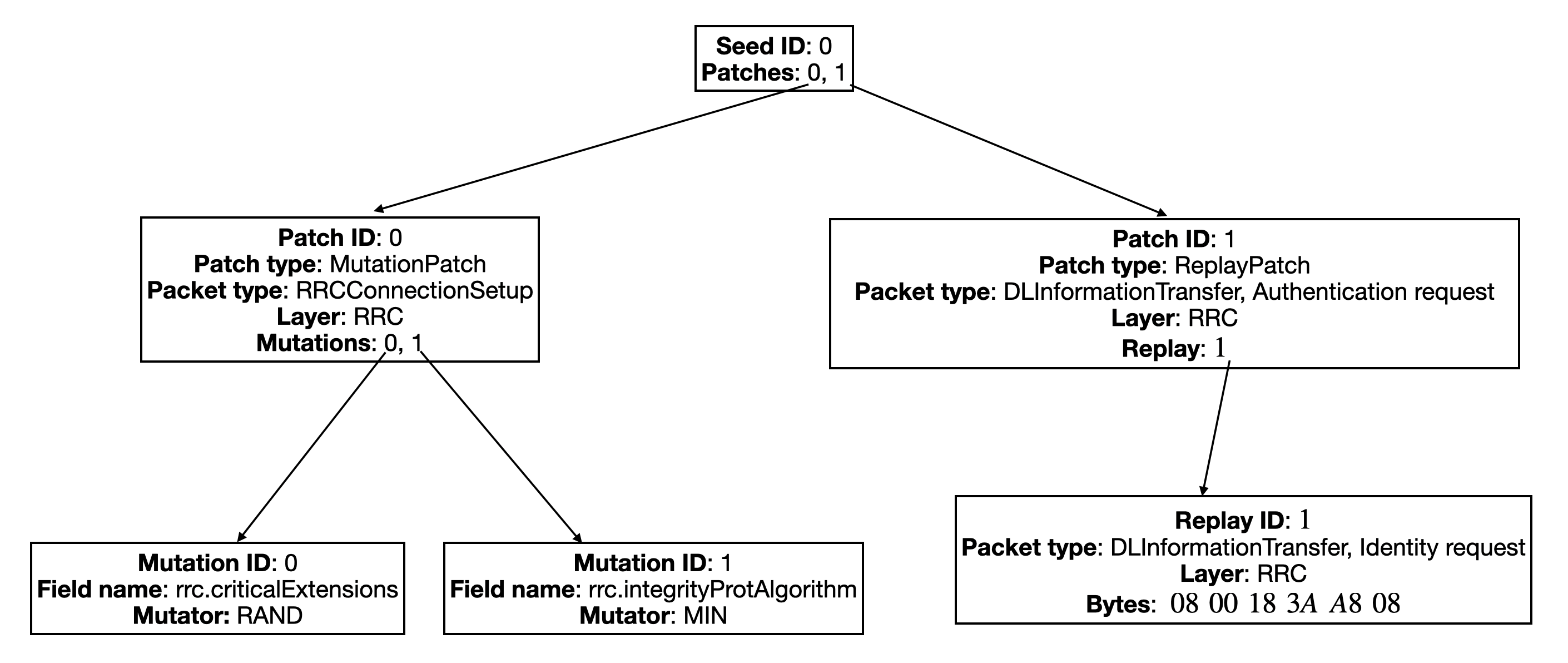}
    \caption{Hierarchical structure of a \textit{Seed}.}
    \label{fig:seed_patch_mutation}
\end{figure}

The \textit{Seed} encapsulates all the \textit{Patches} that have been applied to the \textit{Packets} during a specific fuzzing iteration. Storing the \textit{Seed} enables us to precisely replicate a fuzzing iteration by reapplying all the \textit{Patches} in the exact sequence they were originally applied. This capability ensures reproducibility in our fuzzing experiments.
The aforementioned terms are summarized in Figure~\ref{fig:seed_patch_mutation}.

\section{Fuzzers implemented in the framework} \label{sec:fuzzers}
In this section, we describe the \textit{Fuzzer}s that are implemented in our framework. It should be noted that additional custom fuzzers can be seamlessly integrated into our framework by inheriting from the \textit{Fuzzer} base class.

Before delving into specifics, it is crucial to note that during the \textit{Mutation} phase of our fuzzing process, we adhere to mutation probabilities assigned to each \textit{Field}. These probabilities dictate the likelihood of a particular \textit{Field} being mutated in each fuzzing iteration.
The initial probabilities are assigned uniformly for all the \textit{Field}s in a packet:
\[
    \forall f \in F_P: \ p^0_f = \frac{k}{|F_P|} \text{ where $k$ is a hyper-parameter}
\]
Therefore, initially, the probabilities for the fields are normalized by the number of fields in that packet. To prevent scenarios where a field is either never mutated or always mutated, we constrain the mutation probability value to be between $0.005$ and $0.90$.

The hyper-parameter $k$ determines the average number of fields that will be mutated per packet. This parameter needs careful consideration: if $k$ is too large, the \textit{DUT} will likely reject the packet. Conversely, if $k$ is too small, the fuzzing process may be ineffective. 

\subsection{Random fuzzer} %GC
The \textit{Random fuzzer} implements a straightforward fuzzing strategy. Upon receiving a message to fuzz from the \textit{Fuzzing Controller}, it chooses either a mutation or replay strategy according to the user-defined probabilities.

If the replay strategy is chosen, the intercepted \textit{Packet} is replaced by the old \textit{Packet} from the replay buffer as described in the previous section.

If the mutation strategy is selected, the \textit{Random fuzzer} identifies which \textit{Fields} in the \textit{Packet} will be mutated by evaluating each \textit{Field}'s mutation probability. This process involves iterating through all the \textit{Fields} in the \textit{Packet} and comparing each \textit{Field}'s mutation probability to a randomly generated number between $0$ and $1$. If this random number is less than the \textit{Field}'s mutation probability, the \textit{Field} is designated for mutation. For each selected \textit{Field}, a \textit{Mutation} instance is created using a randomly chosen \textit{Mutator}. Subsequently, the \textit{Random fuzzer} aggregates these \textit{Mutation}s into a \textit{Patch}, which is then applied to the \textit{Packet}.

The \textit{Random fuzzer} operates without any feedback from the \textit{DUT} to optimize the fuzzing process. See Algorithm \ref{alg:random_fuzz} for details.
\begin{algorithm}
    \caption{Implementation of random fuzzing}\label{alg:random_fuzz}
    \begin{algorithmic}[1]
    \State \textbf{Input:} \textit{Packet} \Comment{Intercepted \textit{Packet}}
    \State \textbf{Output:} \textit{Packet} \Comment{\textit{Packet} after fuzzing}
    \LeftComment[0\dimexpr\algorithmicindent]{global float \textit{ReplayProb}} \Comment{User-defined}
    \LeftComment[0\dimexpr\algorithmicindent]{global float \textit{MutProb}} \Comment{User-defined}
    \LeftComment[0\dimexpr\algorithmicindent]{global function \textit{RNG}} \Comment{Random number generator}
    \LeftComment[0\dimexpr\algorithmicindent]{global map \textit{ReplayBuf}} \Comment{Buffer with old \textit{Packets}}
    \State
    \Function{Fuzz}{\textit{Packet}}
        \State \textit{Mutations} $\gets$ \{\} \Comment{Declare a set of \textit{Mutation}s}
        \State \textit{Patch} $\gets$ \{\} \Comment{Declare an empty \textit{Patch}}
        \State $r_1 \gets $\Call{\textit{RNG}}{$ $} \Comment{Get a random number from $0$ to $1$}
        \If {$r_1 <$ \textit{ReplayProb}}
            \State \textit{NewPacket} $\gets$ \Call{\textit{ReplayBuf}.get}{\textit{Packet}.channel}
            \State \Call{\textit{Patch}.addReplay}{\textit{NewPacket}}
        \ElsIf {$r_1 <$ \textit{MutProb}}
            \For {$f$ in \textit{Packet}.fields}
                \State $r_2 \gets $\Call{\textit{\textit{RNG}}}{$ $}
                \If {$r_2 < $ $f$.Prob}
                    \State $m \gets$\Call{CreateMutation}{$f$}
                    \State \Call{\textit{Mutations}.insert}{$m$}
                \EndIf
            \EndFor
            \State \Call{\textit{Patch}.addMutations}{\textit{Mutations}}
        \EndIf
        \State \textbf{return} \Call{\textit{Patch}.apply}{\textit{Packet}}
        \EndFunction
    \end{algorithmic}
\end{algorithm}

\subsection{Coverage-based fuzzer}\label{Coverage-based_fuzzer}
The \textit{Coverage-based fuzzer} is designed to enhance the fuzzing process by maximizing the code coverage of the \textit{DUT} (its 4G/5G implementation to be precise).
This approach draws inspiration from successful modern fuzzers. As the coverage metric, we use edge coverage, which tells us which edges in the program's control flow graph were executed during the fuzzing iteration. Similar to American Fuzzy Lop (AFL)~\cite{AFL}, our fuzzer categorizes edge hits into $8$ buckets based on frequency: $1$, $2$, $3$, $4-7$, $8-15$, $16-31$, $32-127$ and $128+$ times. The main objective of the \textit{Coverage-based fuzzer} is to achieve the highest possible total edge coverage across all iterations, thereby exploring a broad spectrum of execution paths within the \textit{DUT}.

Internally, the \textit{Coverage-based fuzzer} operates similarly to a \textit{Random fuzzer} on a per-iteration basis. However, it introduces a crucial adaptation between iterations: the adjustment of mutation probabilities for \textit{Field}s that were mutated during the previous iteration.
The adjustment equation for mutation probabilities is crafted based on the following principles:
    
\begin{enumerate}[leftmargin=*]\label{list:goals}
     \item If new coverage is discovered in iteration $i$, the mutation probabilities of all mutated fields are increased. Conversely, if no new coverage is found, the corresponding mutation probabilities are decreased.
    
    \item Early in the fuzzing process, the adjustment for new coverage found is less aggressive compared to later stages. This approach acknowledges that discovering new paths in the initial iterations is typically easier than in subsequent ones. Conversely, towards the end of fuzzing, more weight is given to new coverage findings, enhancing exploration in critical areas.
    
    \item The adjustment in mutation probabilities for a field $f$ is moderated by the size of $V_f$, i.e. the set of possible values that $f$ can take. Larger $V_f$ implies more potential values to test, requiring more iterations to ascertain if $f$ is promising for further fuzzing.

\end{enumerate}

The resulting equation for adjusting the probabilities after completing the $i$-th iteration can be seen in Equation~\ref{eq:prob_change}.
\begin{equation}\label{eq:prob_change}
    p^{i}_f \leftarrow p^{i-1}_f + \frac{F(c^{i},i)}{\log_2(|V_f| + 1)}
\end{equation}

The function $F(c^i,i)$ plays a pivotal role in the equation~\ref{eq:prob_change}. It depends on the iteration $i$ and the new coverage $c^i$ found during iteration $i$. The equation for $F(c,i)$ can be seen in Equation~\ref{eq:F},
\begin{equation}\label{eq:F}
    F(c,i) = \frac{ f(c) \cdot g(i)}{n^{i}}
\end{equation}
where $n^i$ represents the total number of mutated fields during iteration $i$ and functions $f(c)$ and $g(i)$ depend on the new coverage found and the fuzzing iteration number, respectively.
Normalization of the probability adjustment is achieved by dividing by $n^{i}$ -- the number of mutated fields.

The equation for $g(i)$, which addresses the second goal mentioned earlier, is depicted in Equation~\ref{eq:g}.
\begin{equation}\label{eq:g}
    g(i) = 
    \begin{cases}
    \beta \cdot \frac{i}{max\_i}, & \text{if}\ c^i > 0 \\
    \frac{1}{\beta \cdot \frac{i}{max\_i}}, & \text{otherwise}
    \end{cases}
\end{equation}
Here, $max\_i$ denotes the total amount of iterations intended by the current fuzzing strategy, and $\beta$ is a user-defined hyper-parameter.  We conducted various experiments to determine an optimal value for $\beta$. The parameter $\beta$ controls the rate at which mutation probabilities are adjusted. Higher values of $\beta$ result in larger increments to the probabilities of mutated fields when new coverage is achieved, while reducing the decrease when coverage remains unchanged. Conversely, lower values of $\beta$ lead to smaller adjustments to mutation probabilities upon new coverage, with larger reductions when coverage stagnates.

For $f(c)$, we selected a function that evaluates to $1$ if new coverage is discovered and $-1$ otherwise as depicted in Equation~\ref{eq:f_1}.

\begin{equation}\label{eq:f_1}
    f(c) =
    \begin{cases}
        1, & \text{if}\ c > 0 \\
        -1, & \text{otherwise}
    \end{cases}
\end{equation}

\section{Evaluation}\label{sec:evaluation}

\subsection{Implementation and setup} % GL
Our framework is implemented in C++. Additionally, we developed patches to srsRAN instances to enable uplink and downlink packet interception. Furthermore, we implemented patches to srsUE, srsEPC and Open5Gs, as described in Section~\ref{protocol_stack_impl}.
All experiments were conducted on an Intel Core i$7$ HP Elitebook $850$ G$8$ with 16GB of RAM and 8 CPU cores, running Ubuntu 22.04 as the operating system.

Our experimental setup consists of two main components. First, we evaluated our grey-box and black-box coverage-based fuzzing algorithms using only srsRAN 4G~\cite{SRSRAN_4G} instances, assessing both uplink and downlink with the \textit{DUT}'s code coverage metrics. Second, we conducted downlink fuzzing experiments targeting COTS 4G and 5G UEs.

\subsection{Fuzzing srsRAN 4G instances} \label{sec:Fuzz_srsRAN}
In our fuzzing experiments on srsRAN 4G instances, we utilized a setup incorporating ZMQ Virtual Radios~\cite{SRSRAN_4G_ZMQ}. Rather than using physical radios, this setup leverages the ZMQ networking library to transfer radio samples between the srsUE and srsENB. All experiments were conducted using srsRAN 4G version 23.11, the latest available version at the time of this research.

In all fuzzing experiments within this subsection, the replay probability was fixed at $0\%$. This decision aimed to focus solely on evaluating the algorithm's ability to adjust field mutation probabilities based on coverage feedback. Additionally, we limited the interception point to the RRC layer, ensuring that only RRC and NAS layer fields were mutated.
Each fuzzing configuration was tested across $20$ sets with $2000$ fuzzing iterations each. It took approximately $1$ hour and $7$ minutes for each set to complete, averaging about two seconds per attach procedure.
The primary metric for evaluating the fuzzing algorithms was the code coverage achieved by the \textit{DUT}. Thus, the effectiveness of a fuzzer was determined by its ability to maximize \textit{DUT}'s code coverage.

Our evaluation is organized as follows: first, we determine the optimal value for the parameter $k$, which represents the average number of fields to be mutated per packet at the start of fuzzing. For the \textit{Random Fuzzer}, the field mutation probabilities remain constant, and the value of $k$ influences the entire fuzzing campaign. Therefore, it is important to select a good value for $k$ for the random fuzzer. Next, we assess the performance of the \textit{Coverage-based Fuzzer} in both grey-box and black-box settings, comparing its results against the optimized \textit{Random Fuzzer}. We also examine how different values of the parameter $\beta$ affect the fuzzer's effectiveness.
Additionally, we establish a ``No Fuzzing" baseline by executing 2000 attach procedures without any fuzzing, using this as the zero-coverage reference point for benchmarking the performance of the fuzzers.

\subsubsection{Estimating parameter k for the \textit{Random fuzzer}} \label{sec:Estimate_k_for_Random_Fuzzer}

\begin{figure}
    \begin{center}
        \resizebox{\columnwidth}{!}{\input{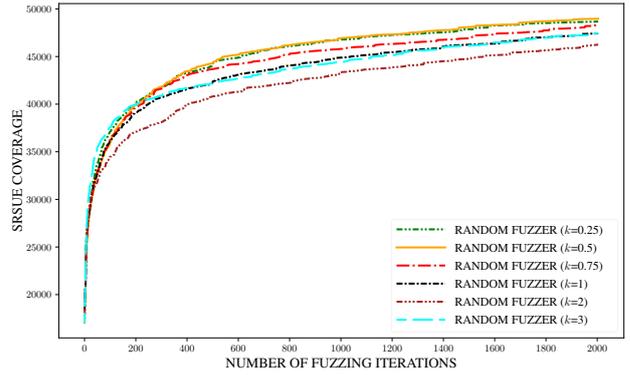}}
    \end{center}
    \vspace{-0.5cm}
    \caption{Downlink random fuzzing with different $k$ parameter values.}
    \label{fig:dl_ks}
\end{figure}

\begin{figure}
    \begin{center}
        \resizebox{\columnwidth}{!}{\input{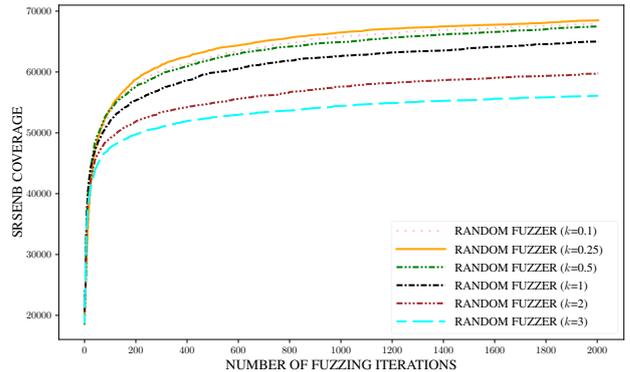}}
    \end{center}
    \vspace{-0.5cm}
    \caption{Uplink random fuzzing with different $k$ parameter values.}
    \label{fig:ul_ks}
\end{figure}

The parameter $k$ is the only adjustable parameter for the \textit{Random fuzzer}. Consequently, we conducted experiments solely using the \textit{Random fuzzer} for this analysis.
The results for downlink and uplink fuzzing are presented in Figures~\ref{fig:dl_ks} and~\ref{fig:ul_ks}, respectively.

In general, smaller $k$ values outperform larger ones in both downlink and uplink fuzzing. For downlink fuzzing, the best performance is achieved with $k=0.5$, while for uplink fuzzing, $k=0.25$ provides the best results. Interestingly, in Figure~\ref{fig:dl_ks} (downlink fuzzing), the larger $k$ values, such as $k=3$, initially perform better during the first $200$ iterations, but their effectiveness degrades over time.

It is important to note that selecting an inappropriate $k$ value can significantly reduce performance, particularly in uplink fuzzing. The reason for the differing performance of the same $k$ values in downlink versus uplink scenarios can be explained by Table~\ref{tab:DL_UL_STATE_COMP}, which shows the ``length" of each fuzzing iteration, quantified by the average number of packets per iteration. The table reveals that the number of packets is consistently lower in the uplink scenario, indicating that the base station and core network are more hesitant to respond to fuzzed messages than the UE.

\begin{table}[!h]
\caption{Average number of packets sent during the fuzzing depending on $k$ value for uplink and downlink random fuzzing.}\label{tab:DL_UL_STATE_COMP}
\begin{center}
\begin{tabular}{|c|c|c|c|c|c|}
\hline
 \textbf{k}  & \textbf{0.25} & \textbf{0.5}  & \textbf{1}   & \textbf{2}   & \textbf{3}   \\ \hline
\textbf{Downlink} & $13.0$ & $10.6$ & $7.6$ & $5.2$ & $4.6$ \\ \hline
\textbf{Uplink} & $10.5$ & $7.3$  & $4.2$ & $2.1$ & $1.4$ \\ \hline
\end{tabular}
\end{center}
\end{table}

In the following sections, when comparing fuzzers, we consider the \textit{Random Fuzzer} only with the optimal parameters: $k=0.5$ for downlink fuzzing and $k=0.25$ for uplink fuzzing.

\subsubsection{Evaluating our \textit{Coverage-based fuzzer} in the grey-box scenario}
First, we evaluate the performance of the \textit{Coverage-based Fuzzer} when access to coverage information from the \textit{DUT} is available. This scenario is particularly relevant for manufacturers with access to the \textit{DUT}’s source code or in emulation-based fuzzing setups where coverage data can be extracted from the emulated \textit{DUT}.

As discussed in Section~\ref{Coverage-based_fuzzer}, a key hyperparameter in our coverage-based algorithm is $\beta$, which governs the rate at which mutation probabilities are adjusted. We aim to empirically identify the most effective values of $\beta$ for both downlink and uplink fuzzing. In addition, we aim to compare the performance of our coverage-based algorithm with that of the optimized \textit{Random Fuzzer}.

\begin{figure}[!h]
    \begin{center}
        \resizebox{\columnwidth}{!}{\input{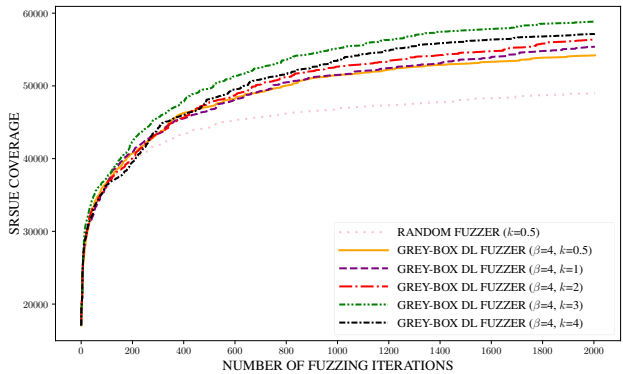}}
    \end{center}
    \vspace{-0.5cm}
    \caption{Grey-box downlink fuzzing with different $k$ parameter values.}
    \label{fig:greybox_dl_ks}
\end{figure}

\begin{figure}[!h]
    \begin{center}
        \resizebox{\columnwidth}{!}{\input{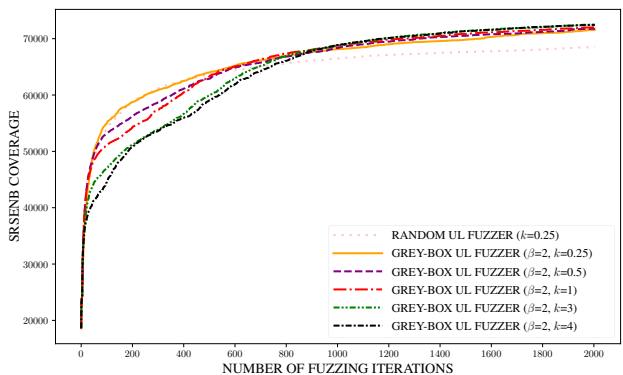}}
    \end{center}
    \vspace{-0.5cm}
    \caption{Grey-box uplink fuzzing with different $k$ parameter values.}
    \label{fig:greybox_ul_ks}
\end{figure}

In our experiments for the \textit{Coverage-based fuzzer}, we did not restrict ourselves to the $k$ values that performed best with random fuzzing but also explored other values. Our empirical findings (see Figures~\ref{fig:greybox_dl_ks} and~\ref{fig:greybox_ul_ks}) revealed that using $k=3$ in conjunction with the \textit{Coverage-based Fuzzer} provided the best results, even though in case of uplink fuzzing, the differences in performance were minor. Therefore, in this section, we exclusively use $k=3$ for coverage-based fuzzing.

\begin{figure}
    \begin{center}
        \resizebox{\columnwidth}{!}{\input{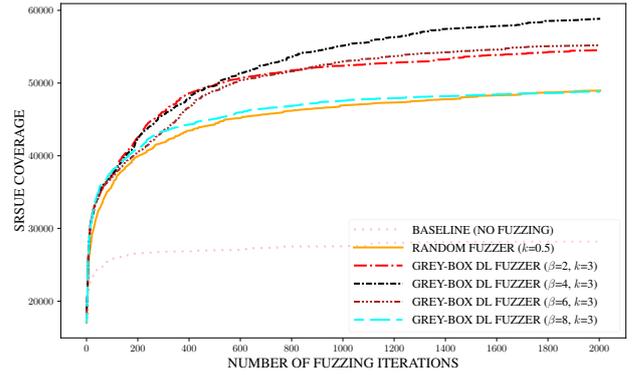}}
    \end{center}
    \vspace{-0.5cm}
    \caption{Grey-box downlink fuzzing with different $\beta$ parameter values.}
    \label{fig:dl_betas}
\end{figure}

\textbf{Downlink fuzzing.}
The results of the grey-box \textit{Coverage-based fuzzer}s with different $\beta$ values are presented in Figure~\ref{fig:dl_betas}. For comparison, the figure also includes the performance of the best downlink \textit{Random Fuzzer} as well as the ``No Fuzzing'' baseline.

We experimented with $\beta$ values ranging from $1$ to $9$ but, for clarity, only display the results for $\beta$ values of $2$, $4$, $6$, and $8$ in the figure.
Several observations can be made based on these results:
\begin{itemize}
    \item \textbf{Impact of $\beta$ value}: The graph demonstrates that the choice of $\beta$ significantly influences the code coverage achieved by the \textit{DUT}. Among the tested values, $\beta=4$ proves to be the most effective for grey-box downlink fuzzing.
    \item \textbf{Comparison with \textit{Random Fuzzer}}: The grey-box \textit{Coverage-based Fuzzer}, particularly with $\beta=4$, outperforms the optimized \textit{Random Fuzzer} substantially, achieving a 47.6\% increase in code coverage compared to the \textit{Random Fuzzer}\footnotemark[2].
\end{itemize}

\footnotetext[2]{Code coverage increase is measured relative to the ``No Fuzzing" baseline, which is treated as having 0 coverage.}

\begin{figure}
    \begin{center}
        \resizebox{\columnwidth}{!}{\input{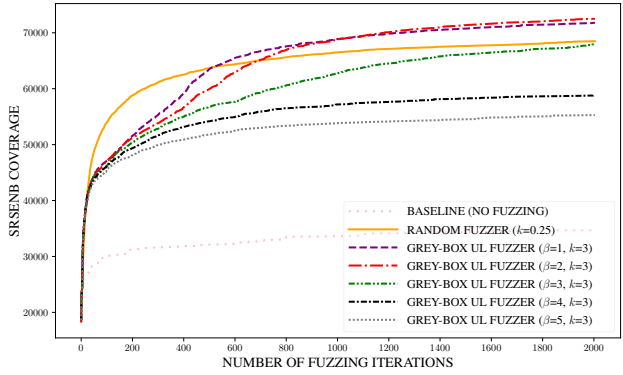}}
    \end{center}
    \vspace{-0.5cm}
    \caption{Grey-box uplink fuzzing with different $\beta$ parameter values.}
    \label{fig:ul_betas}
\end{figure}

\textbf{Uplink fuzzing.}
In the uplink fuzzing scenario, we conducted experiments similar to those in the downlink case, with the \textit{DUT} being the srsENB. We evaluated the \textit{Random Fuzzer}, \textit{Coverage-based Fuzzers} with varying $\beta$ values, the optimized \textit{Random Fuzzer}, and a ``No Fuzzing" baseline. Notably, larger $\beta$ values did not perform well in this scenario, so we limited our experiments to $\beta$ values between $1$ and $5$. The results are summarized in Figure~\ref{fig:ul_betas}. The key observations from these experiments are:
\begin{itemize}
    \item \textbf{Impact of choosing k=3}:
    Since we set $k=3$, the fuzzing initially behaves similarly to the \textit{Random Fuzzer} with the same parameter (see Figure~\ref{fig:ul_ks}), w, which underperforms compared to smaller $k$ values. However, due to the coverage-based algorithm, the performance improves rapidly, with the fuzzer surpassing the \textit{Random Fuzzer} after approximately 500 iterations. Had we used $k=0.25$, the initial performance would have been better, but over time, it would have been slightly outperformed by the fuzzer using $k=3$ (see Figure~\ref{fig:greybox_ul_ks}).
    \item \textbf{Impact of $\beta$ value}: Similar to the downlink scenario, varying the $\beta$ parameter significantly affects the code coverage achieved. In the uplink case, the best performance was observed with $\beta=2$. Interestingly, smaller $\beta$ values were more effective in uplink fuzzing than in downlink, likely due to the larger size of the srsENB software compared to srsUE. As a result, coverage increases were more frequent, and larger $\beta$ values led to an explosion of mutation probabilities, reducing the effectiveness of fuzzing (see Section~\ref{app:mutations} for details).
    \item \textbf{Comparison with \textit{Random Fuzzer}}: In the uplink scenario, the grey-box \textit{Coverage-based Fuzzer} with smaller $\beta$ values outperformed the optimized \textit{Random Fuzzer}. Specifically, with $\beta=2$, the \textit{Coverage-based Fuzzer} achieved an $11.9\%$ increase in code coverage compared to the \textit{Random Fuzzer}.
\end{itemize}

To summarize, the \textit{Coverage-based fuzzer}s in grey-box settings allows to increase the \textit{DUT}'s coverage by $47.6\%$ in the downlink and by $11.9\%$ in the uplink scenarios over the optimized \textit{Random fuzzer}s. However, the performance of the \textit{Coverage-based fuzzer} depends on carefully selecting the $\beta$ parameter, which is crucial in optimizing its effectiveness. More detailed insights into how the parameter $\beta$ influences the number of mutations and subsequently the coverage can be found in Section~\ref{app:mutations}.

\subsubsection{\textbf{Evaluating our \textit{Coverage-based} algorithm in the black-box scenario}}
In the black-box scenario, where direct access to the DUT's coverage information is unavailable, we propose leveraging code coverage metrics obtained from the srsENB for downlink fuzzing and the srsUE for uplink fuzzing. The underlying hypothesis is that there exists a strong correlation between the code coverage of the UE and the base station due to their constant communication during the 4G/5G protocol execution.
To test this hypothesis, we estimated the optimal $\beta$ parameter for the black-box scenario and compared the results with both the \textit{Random Fuzzer} and the grey-box coverage-based scenario. Like the grey-box case, we observed that $k=3$ yields the best performance. Therefore, all black-box experiments were conducted using this value for the $k$ parameter.

\textbf{Downlink fuzzing.}
In the downlink fuzzing scenario, the \textit{DUT} is again the srsUE. However, as mentioned previously, we do not utilize the coverage information from srsUE in our coverage-based algorithm. Instead, we rely on the code coverage metrics obtained from the srsENB. The coverage information from the srsUE is only collected to evaluate and compare the fuzzers' effectiveness.

\begin{figure}
    \begin{center}
        \resizebox{\columnwidth}{!}{\input{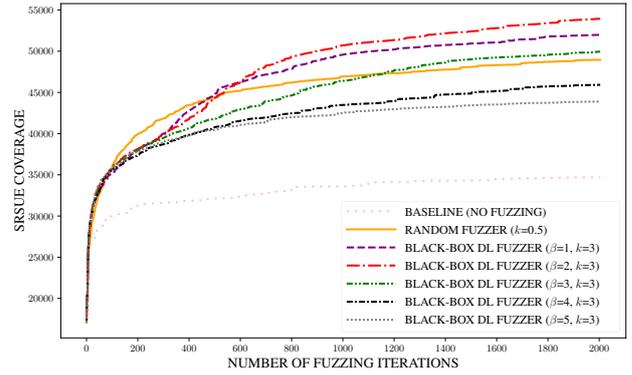}}
    \end{center}
    \vspace{-0.5cm}
    \caption{Black-box downlink fuzzing with different $\beta$ parameter values.}
    \label{fig:dl_black_box}
\end{figure}

Similar to the grey-box uplink scenario, we ran experiments with $\beta$ values ranging from $1$ to $5$, as larger $\beta$ values do not perform well when the base station guides the fuzzing. The results are shown in Figure~\ref{fig:dl_black_box}. As observed, smaller $\beta$ values are again preferred for optimal performance. Specifically, we found that $\beta=2$ yields the best code coverage in this scenario.

Furthermore, while directly obtaining the coverage from srsUE (as done in the grey-box downlink fuzzing) resulted in better coverage, the \textit{Coverage-based fuzzer} guided by coverage feedback from srsENB still achieves a $23.9\%$ increase in code coverage over the \textit{Random fuzzer}. This demonstrates that even without direct access to the \textit{DUT}'s coverage information, our approach can still significantly improve fuzzing effectiveness.

\begin{figure}
    \begin{center}
        \resizebox{\columnwidth}{!}{\input{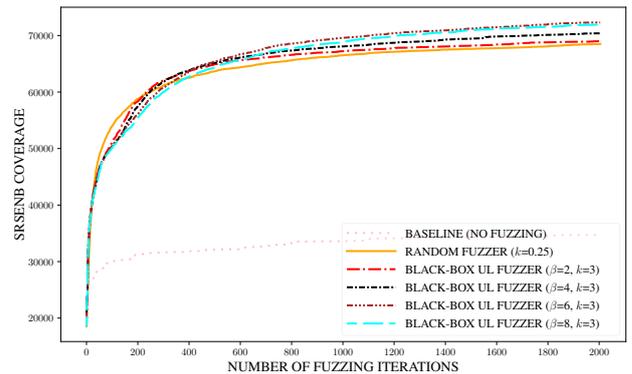}}
    \end{center}
    \vspace{-0.5cm}
    \caption{Black-box uplink fuzzing with different $\beta$ parameter values.}
    \label{fig:ul_black_box}
\end{figure}

\textbf{Uplink fuzzing.}
Similarly, we conducted experiments for uplink fuzzing, where the \textit{DUT} is the srsENB, using coverage feedback from the srsUE. The results of these experiments are depicted in Figure~\ref{fig:ul_black_box}, which shows the fuzzer's performance with different $\beta$ values (graphs for $\beta=2$, $4$, $6$, and $8$ are presented). Unlike in the previous scenarios, the choice of $\beta$ has a smaller impact on the performance of the fuzzer in the uplink case, with most of the graphs being relatively close to each other. Nevertheless, we observe that $\beta=6$ yields the best performance.
Interestingly, the best-performing black-box uplink \textit{Coverage-based fuzzer} performs similarly to the grey-box version, yielding an increase of $11.3\%$ in the code coverage over the \textit{Random fuzzer}.

\subsection{Average number of mutations per packet} \label{app:mutations}

In order to get a deeper insight into the hyperparameters, we analyze the graphs that display the average number of mutations per packet during the fuzzing of srsRAN 4G instances. Our primary objective is to explore the correlation between these mutation patterns and the coverage graphs presented in Section~\ref{sec:Fuzz_srsRAN}.

While the number of mutations provides useful insight into the fuzzing algorithm, it cannot be the sole indicator of performance. This is because the probabilities of mutating each field are adjusted independently, aiming to increase the probabilities of promising fields while decreasing those of less relevant ones. Consequently, two fuzzers may exhibit similar mutation rates per packet but produce different coverage outcomes.

\subsubsection{Downlink fuzzing}
\begin{figure}[]
    \begin{center}
        \resizebox{\columnwidth}{!}{\input{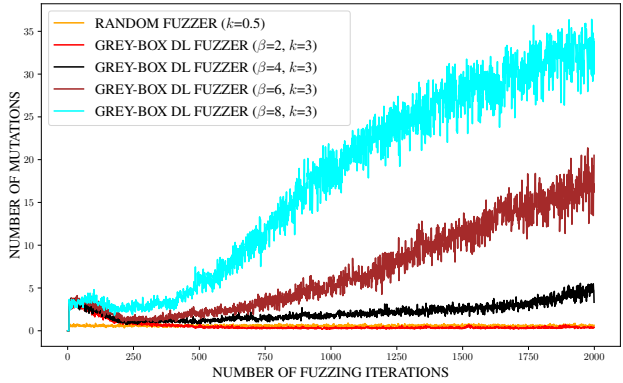}}
    \end{center}
    \vspace{-0.5cm}
    \caption{Average number of mutations per packet during the grey-box downlink fuzzing with different $\beta$ parameter values. Higher $\beta$ values lead to the explosion of the probabilities and hence inefficient fuzzing.}
    \label{fig:greybox_dl_betas_mut}
\end{figure}

\begin{figure}[]
    \begin{center}
        \resizebox{\columnwidth}{!}{\input{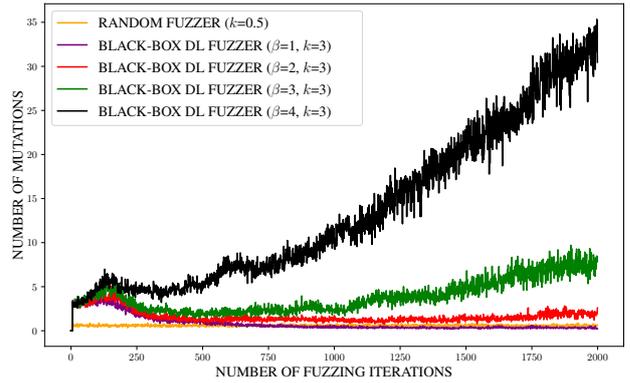}}
    \end{center}
    \vspace{-0.5cm}
    \caption{Average number of mutations per packet during the black-box downlink fuzzing with different $\beta$ parameter values. Higher $\beta$ values lead to the explosion of the probabilities and hence inefficient fuzzing.}
    \label{fig:blackbox_dl_betas_mut}
\end{figure}

Figures~\ref{fig:greybox_dl_betas_mut} and~\ref{fig:blackbox_dl_betas_mut} depict the results for grey-box and black-box downlink fuzzing, respectively (to maintain a consistent scale between the two, we omitted the mutation graph for black-box fuzzing with $\beta=5$). These mutation graphs can be compared to the respective coverage graphs shown in Figures~\ref{fig:dl_betas} and~\ref{fig:dl_black_box}. 
First, as expected, the \textit{Random fuzzer} with $k=0.5$ maintains an average number of mutated fields per packet close to $0.5$ throughout the entire fuzzing campaign.
Second, for the \textit{Coverage-based fuzzer}, the larger the $\beta$ value, the more fields are mutated in the long run. From the coverage graphs, we observed that larger $\beta$ values do not perform well, particularly in the latter half of the fuzzing campaign. This performance decline is caused by the explosion of mutation probabilities, leading to an excessive number of mutated fields per iteration, as illustrated by Figures~\ref{fig:greybox_dl_betas_mut} and~\ref{fig:blackbox_dl_betas_mut}. 
Notably, the $\beta$ values that delivered the best coverage -- $\beta=4$ for grey-box and $\beta=2$ for black-box downlink fuzzing -- managed to strike a good balance between the explosion of probabilities and their fading.

\subsubsection{Uplink fuzzing}
\begin{figure}[]
    \begin{center}
        \resizebox{\columnwidth}{!}{\input{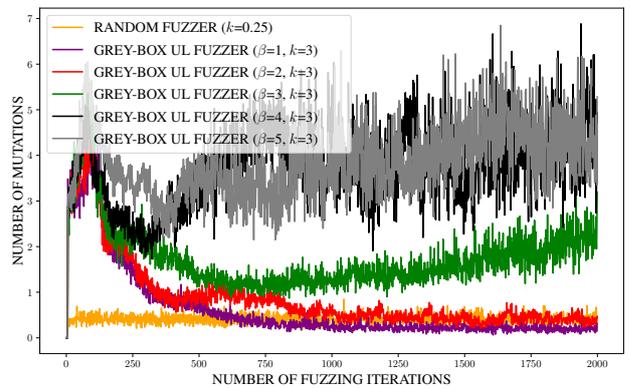}}
    \end{center}
    \vspace{-0.5cm}
    \caption{Average number of mutations per packet during the grey-box uplink fuzzing with different $\beta$ parameter values.}
    \label{fig:greybox_ul_betas_mut}
\end{figure}

\begin{figure}[!h]
    \begin{center}
        \resizebox{\columnwidth}{!}{\input{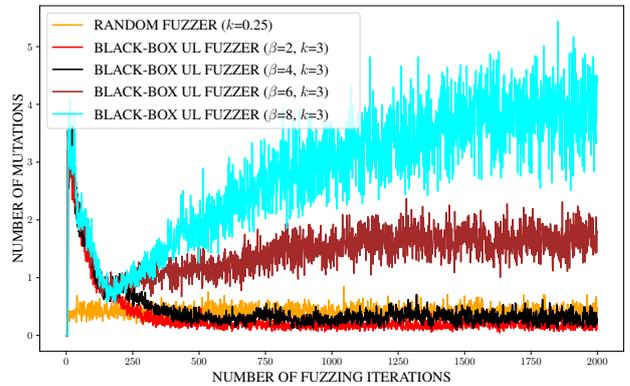}}
    \end{center}
    \vspace{-0.5cm}
    \caption{Average number of mutations per packet during the black-box uplink fuzzing with different $\beta$ parameter values.}
    \label{fig:blackbox_ul_betas_mut}
\end{figure}

For the uplink scenario, Figures~\ref{fig:greybox_ul_betas_mut} and~\ref{fig:blackbox_ul_betas_mut} show the mutation results for grey-box and black-box fuzzing, respectively, and can be compared with the corresponding coverage graphs in Figures~\ref{fig:ul_betas} and~\ref{fig:ul_black_box}.
Overall, the same trends observed in the downlink scenario are evident here. Larger $\beta$ values still lead to an increase in the number of mutated fields, though the probability explosion is less severe compared to the downlink fuzzing. However, as discussed in Section~\ref{sec:Estimate_k_for_Random_Fuzzer}, the base station and core network are more reluctant to proceed with the attach procedure after receiving a mutated packet, unlike the UE. Thus, even a modest increase in mutation probabilities can result in stagnating coverage progress.

\hfill \break
To conclude, this section provided an analysis of the hyperparameters in our fuzzing framework and their influence on overall fuzzing effectiveness. We demonstrated that incorrect hyperparameter choices for any fuzzer can result in suboptimal performance. The best-performing parameters for each scenario are outlined in Table~\ref{tab:best_params}.
Furthermore, our results confirm that the proposed coverage-based algorithm outperforms the optimized \textit{Random fuzzer} in both grey- and black-box settings.

\begin{table}[!h] 
\centering
\caption{Summary of the best hyperparameter values}
\label{tab:best_params}
\begin{tabular}{|c|c|c|}
\hline
                   & \textbf{Downlink} & \textbf{Uplink} \\ \hline
\textbf{Random fuzzer}    &      $k=0.5$       &      $k=0.25$       \\ \hline
\textbf{Grey-box fuzzer}  &       $k=3, \beta=4$      &      $k=3, \beta=2$       \\ \hline
\textbf{Black-box fuzzer} &     $k=3, \beta=2$        &      $k=3, \beta=6$      \\ \hline
\end{tabular}
\end{table}

\section{Fuzzing COTS 4G/5G UEs}\label{Fuzzing_COTS_UEs}
We acquired multiple 4G/5G COTS UEs to evaluate our framework on real devices. All 5G modems also support 4G, and thus we conducted fuzzing tests in both 4G and 5G modes. These UEs mostly consisted of bare 4G/5G modules with mPCIe or M.2 pinouts. To establish communication with the modems, we utilized corresponding USB adapters. Once connected to a PC, these modems expose a virtual serial interface through which AT commands can be sent to communicate with the modem. However, there were two exceptions, namely the Nokia Gateway~\cite{Nokia_5G21} and Samsung Galaxy A7~\cite{Samsung_A7} android smartphone, both of which did not provide a direct access to the AT commands.
For the Nokia Gateway, we gained access to a root shell on the device, allowing us to send AT commands directly to the underlying 5G modem. In the case of the Samsung Galaxy A7, we used ADB to toggle airplane mode. To prevent the fuzzed packets from being transmitted unintentionally to other devices, the smartphone was placed in a RF enclosure~\cite{jre-enclosure}. As the enclosure is not transparent, scrcpy~\cite{scrcpy} was a useful tool for viewing the screen of the Android device from the PC.

Since COTS UEs do not readily provide a method to collect coverage information, we utilized the black-box mode of our coverage-based fuzzer. This approach guided the fuzzing process using coverage information obtained from srsENB or srsGNB, depending on the target device. For each target and supported protocol, we conducted 10 fuzzing sprints, with each sprint consisting of 5000 iterations (equivalent to attach attempts by the UE) resulting in around 70 hours of testing per device.
Table~\ref{tab:commercial_ues} presents an overview of the COTS 4G/5G UEs tested, alongside key findings. The firmware versions of the devices were acquired with the help of QCSuper~\cite{QCSuper} when possible. In the other cases, the appropriate AT command, namely ATI or AT+CMGR was used. For the smartphone, the firmware version of the baseband was taken from its settings.

Among the $12$ devices tested, one 4G modem and one 5G device experienced connectivity issues with our respective networks, despite 5G device successfully connecting to our 4G network. Similar connection issues were reported by Khandker et al.~\cite{ASTRA-5G} when testing smartphones. Notably, in $10$ devices, the vulnerabilities leading to modem crashes and subsequent reboots were identified. Interestingly enough, all of the tested 5G devices proved to be vulnerable during the 4G attach procedure.

\begin{table}[!b]
\caption{List of tested COTS UEs}\label{tab:commercial_ues}
\begin{center}
\resizebox{\columnwidth}{!}{%
\begin{tabular}{|c|c|c|c|}
\hline
\textbf{UE model}   &
\textbf{Firmware version} &
\textbf{4G}  & \textbf{5G} \\
\hline        
SIMCom SIM7600E~\cite{SIMCom_SIM7600E} & MPSS.JO.2.0.1.c1\-00188\-9607
&  \cellcolor{yellow!75}\textbf{NC} & \cellcolor{black!15}\textbf{-}  \\
\hline
Fibocom L610-EU~\cite{Fibocom_L610EU}  & 16000.1000.00.88.03.02 
&  \cellcolor{red!50}\textbf{V}  & \cellcolor{black!15}\textbf{-}  \\
\hline
Quectel EC25-E~\cite{Quectel_EC25}   &    EC25EFAR06A11M4G   &  \cellcolor{red!50}\textbf{V} & \cellcolor{black!15}\textbf{-}  \\
\hline
Sierra Wireless WP7607~\cite{SW_WP7607} & 
MPSS.JO.2.0.2.c1\-00080\-9607
& \cellcolor{red!50}\textbf{V}  & \cellcolor{black!15}\textbf{-}  \\
\hline
Telit LE910C1-EU~\cite{Telit_LE910C1} & 25.21.220  &  \cellcolor{red!50}\textbf{V}  & \cellcolor{black!15}\textbf{-}  \\
\hline
ZTEWelink ME3630~\cite{ZTEWelink_ME3630} & ME3630E1CV1.0B06     &  \cellcolor{red!50}\textbf{V}  & \cellcolor{black!15}\textbf{-}  \\
\hline
Samsung Galaxy A7~\cite{Samsung_A7} &  A750FNXXU5CVG1 &  \cellcolor{green!50}\textbf{T}  & \cellcolor{black!15}\textbf{-}  \\
\hline
Fibocom FM160-EAU~\cite{Fibocom_FM160EAU} &
89611.1000.00.04.01.07 &  \cellcolor{red!50}\textbf{V}  & \cellcolor{green!50}\textbf{T}   \\
\hline
Sierra Wireless EM9291~\cite{SW_EM9291} & 
 MPSS.DE.3.0.c2-00061-OLYMPIC & \cellcolor{red!50}\textbf{V}  & \cellcolor{green!50}\textbf{T}   \\
\hline
Quectel RM500Q-GL~\cite{QUECTEL_RM500Q} &
MPSS.HI.2.0.c3-00246-SDX55 & \cellcolor{red!50}\textbf{V}  & \cellcolor{red!50}\textbf{V}   \\
\hline
SIMCom SIM8260G-M2~\cite{SIMCom_8380G} & MPSS.DE.4.0.c1-00105-OLYMPIC & \cellcolor{red!50}\textbf{V}  & \cellcolor{green!50}\textbf{T}   \\
\hline
Nokia 5G21 Gateway~\cite{Nokia_5G21}  & MPSS.HI.2.0.c3-00380-SDX55   &  \cellcolor{red!50}\textbf{V}  & \cellcolor{yellow!75}\textbf{NC}   \\
\hline
\multicolumn{4}{l}{ } \\
\multicolumn{4}{l}{\hspace{1cm}\textbf{-} - Protocol is not supported by the UE} \\
\multicolumn{4}{l}{\hspace{1cm}\textbf{NC} - No Connection (we could not connect UE to our network)} \\
\multicolumn{4}{l}{\hspace{1cm}\textbf{V} - Vulnerabilities were found during the fuzzing tests} \\
\multicolumn{4}{l}{\hspace{1cm}\textbf{T} - Tested and no vulnerabilities were found} \\
\end{tabular}
}

\end{center}
\end{table}

\subsection{Discovered vulnerabilities}
During the evaluation of our fuzzing framework, we identified the vulnerabilities in $10$ commercial devices as well as in all of the srsRAN 4G instances (srsUE, srsENB and srsEPC). All findings were promptly reported to the respective manufacturers.

All the found vulnerabilities are summarized in Table~\ref{tab:results}, where we specify the model of the vulnerable device, the layer and the protocol affected, the packet type and the manipulation required to trigger the vulnerability, observed outcome, and any known reasons for the vulnerability. The same values ``Manipulation of the packet" column mean that the vulnerability-triggering seed is identical. The exact seeds that trigger the vulnerabilities are available alongside the code of our fuzzing framework.

\begin{table*}[htbp]
\caption{Fuzzing results}\label{tab:results}
\small
% \begin{center}
\resizebox{\textwidth}{!}{
\begin{tabular}{|c|c|c|c|c|c|c|}
\hline
\textbf{DUT} & \textbf{Protocol} & \textbf{Layer}  & \textbf{Packet type} & \textbf{Manipulation of the packet} & \textbf{Result} & \textbf{Reason} \\
\hline
Sierra Wireless WP7607 & 4G & RRC & \textit{One of the NAS messages}\footnotemark[3] & Fuzz \textit{lte-rrc.dedicatedInfoType} field & Crash & Hit assert\footnotemark[4]\\
\hline
Quectel EC25-E & 4G & RRC & \textit{RRCConnectionSetup} & Fuzz \textit{lte-rrc.sr\_PUCCH\_ResourceIndex} field & Crash & Unknown \\
\hline
Fibocom L610-EU & 4G & RRC & \textit{RRCConnectionSetup} & Fuzz \textit{lte-rrc.sr\_PUCCH\_ResourceIndex} field & Crash & Unknown \\
\hline
Sierra Wireless WP7607 & 4G & RRC & \textit{RRCConnectionSetup} & Fuzz \textit{lte-rrc.sr\_PUCCH\_ResourceIndex} field & Crash & Exception \\
\hline
Telit LE910C1-EU & 4G & RRC & \textit{RRCConnectionSetup} & Fuzz \textit{lte-rrc.sr\_PUCCH\_ResourceIndex} field & Crash & Unknown \\
\hline
ZTEWelink ME3630 & 4G & RRC & \textit{RRCConnectionSetup} & Fuzz \textit{lte-rrc.sr\_PUCCH\_ResourceIndex} field & Crash & Unknown \\
\hline
Fibocom FM160-EAU & 4G & RRC & \textit{RRCConnectionSetup} & Fuzz \textit{lte-rrc.sr\_PUCCH\_ResourceIndex} field & Crash & Unknown \\
\hline
Sierra Wireless EM9291 & 4G & RRC & \textit{RRCConnectionSetup} & Fuzz \textit{lte-rrc.sr\_PUCCH\_ResourceIndex} field & Crash & Exception \\
\hline
SIMCom SIM8260G-M2 & 4G & RRC & \textit{RRCConnectionSetup} & Fuzz \textit{lte-rrc.sr\_PUCCH\_ResourceIndex} field & Crash & Unknown \\
\hline
Nokia 5G21 Gateway & 4G & RRC & \textit{RRCConnectionSetup} & Fuzz \textit{lte-rrc.sr\_PUCCH\_ResourceIndex} field & Crash & Unknown \\
\hline
Quectel RM500Q-GL & 4G & RRC & \textit{RRCConnectionSetup} & Fuzz \textit{lte-rrc.sr\_PUCCH\_ResourceIndex} field & Crash & Unknown \\
\hline
Quectel RM500Q-GL & 5G & RRC & \textit{RRC Setup} & Fuzz \textit{nr-rrc.monitoringSymbolsWithinSlot} field & Crash & Unknown \\
\hline
Quectel RM500Q-GL & 5G & RRC & \textit{RRC Setup} & Fuzz \textit{nr-rrc.tag\_Id} field & Crash & Unknown \\
\hline
Quectel RM500Q-GL & 5G & RRC & \textit{RRC Setup} & Fuzz \textit{nr-rrc.CSI\_IM\_ResourceId} field & Crash & Unknown \\
\hline
Quectel RM500Q-GL & 5G & RRC & \textit{RRC Setup} & Fuzz \textit{nr-rrc.srs\_ResourceSetId} field & Crash & Unknown \\
\hline
srsEPC & 4G & RRC           &  \textit{ULInformationTransfer,IdentityResponse}     &    Send the same IdentityResponse message twice    &   Crash & Use-after-free~\cite{BUG_RRC_ULInfTransfer_IdResp_Inj} \\
\hline

srsEPC & 4G & RRC    & \textit{ULInformationTransfer,IdentityResponse} & Fuzz \textit{per.choice\_index} field & Crash & Buffer Overflow~\cite{BUG_RRC_ULInfTransfer_IdResp_Mut} \\
\hline

srsENB & 4G & RRC           &       \textit{RRCConnectionReconfigurationComplete}      &   Send RRCConnectionReconfigurationComplete out-of-order\footnotemark[5]   &  Crash & Buffer Overflow~\cite{BUG_RRC_ConnReconfigComplete} \\
\hline

srsENB & 4G & RRC        &      \textit{RRCConnectionRequest}       &   Fuzz \textit{m\_TMSI} and \textit{per.choice\_index} fields &     Crash  &  Unknown~\cite{BUG_RRC_Conn_Request} \\
\hline

srsUE & 4G & RRC         &      \textit{RRCConnctionSetup}       &     Fuzz \textit{cqi\_ReportConfig\_element} field    &    Hang  &  Unknown~\cite{BUG_RRC_ConnSetup} \\
\hline

srsUE & 4G & RRC         &       \textit{SecurityModeCommand}      &     Fuzz \textit{securityAlgorithmConfig\_element} and  \textit{integrityProtAlgorithm} fields   & Crash & Buffer Overflow~\cite{BUG_RRC_SecModComm} \\
\hline
srsUE & 4G & RRC     &      \textit{DLInformationTransfer,SecurityModeCommand}       &   Fuzz \textit{nas\_eps.emm.toi} field    &  Crash & Buffer Overflow~\cite{BUG_RRC_DLInfTransfer_SecModeComm} \\
\hline
\end{tabular}
}
% \end{center}
\end{table*}

\afterpage{\footnotetext[3]{Authentication Request or Identity Request or Security Mode Command}
\footnotetext[4]{This vulnerability was already known to the manufacturer when we reported it (CVE-2022-40504~\cite{CVE-2022-40504})}
\footnotetext[5]{In order to trigger the crash in srsENB an attcker should send a RRCConnectionReconfigurationComplete message instead of the one of the following messages: Identity Request, Authentication Request, Security mode command or UE Capability Enquiry}
}
\subsubsection{Vulnerabilities in srsRAN 4G instances}
Among the $7$ vulnerabilities discovered in srsRAN 4G instances, $6$ resulted in crashes of various srsRAN components, while $1$ caused srsUE to hang. The most prevalent type of vulnerability was buffer overflow, occurring $4$ times. Notably, $2$ of these buffer overflows occurred during logging activities. The use of address sanitizer facilitated detecting the buffer-overflow and use-after-free vulnerabilities in the srsRAN software.

 \subsubsection{Vulnerabilities in the COTS UEs}
In embedded devices, common vulnerabilities such as buffer overflow or use-after-free do not always lead to an immediate crash~\cite{What_You_Corrupt_Is_Not_What_You_Crash}. Even when a crash does occur, identifying its exact root cause can be difficult, particularly in the absence of debugging information. Some manufacturers attempt to alleviate this problem by adding specific sets of AT commands that provide debugging data after a modem reboots. However, among the devices we tested, this feature was only present in two Sierra Wireless modems. As a result, for most of the tested commercial devices, we were unable to determine the precise reason for their crashes.

In our efforts to reproduce crashes, we employed two methods, focusing on the last seed used before the crash, which we call the \textit{crash seed}.
The first method involved fully replaying the \textit{crash seed}, repeating all replay and mutation patches exactly as they were applied during the original fuzzing session. However, we noticed that this approach sometimes failed to consistently reproduce the crash.
In response, our second approach treated all patches as replay patches. Instead of mutating individual fields of a new packet, we replayed the entire packet from the \textit{crash seed}. This modification proved more reliable, successfully reproducing the majority of crashes encountered during our fuzzing efforts.

However, sometimes we have encountered crashes that we could not reproduce. These cases are not included in Table~\ref{tab:results}. Determining the exact cause of those crashes is challenging, but they could potentially be caused by the memory-safety vulnerabilities triggered in earlier fuzzing iterations. This phenomenon of delayed crashes was also mentioned in~\cite{What_You_Corrupt_Is_Not_What_You_Crash}.

For the crashes listed in Table~\ref{tab:results}, there were instances where multiple payloads could trigger the same crash. When it was unclear whether these were distinct crashes, we opted not to differentiate them in the table.

\subsection{Attack scenario \& Impact of found vulnerabilities}

\subsubsection{Downlink fuzzing}
Assume the vulnerability in the UE is triggered upon receiving malicious packet \textit{X}.
An attacker could exploit this vulnerability as follows:
\begin{enumerate}
    \item An attacker sets up a malicious base station.
    \item UE tries to connect to the malicious base station and triggers an attach procedure.
    \item During the attach procedure, the malicious base station sends malicious packet \textit{X} to the UE, triggering the vulnerability.
\end{enumerate}

The second step can occur due to various reasons. For example, when the modem is initially disconnected from any network (such as after a flight), it performs a full scan of available networks. Upon detecting the malicious base station, the modem attempts to connect to it. Alternatively, a user might manually select the malicious network for connection, during which the malicious message is sent.

When evaluating the impact of the vulnerabilities a couple of factors should be considered. Firstly, it matters whether message \textit{X} is sent before or after the network is authenticated by the UE. In the case where \textit{X} is sent after authentication, the exploit would require the SIM card to successfully authenticate the network before the base station can send message \textit{X} to trigger the vulnerability. This could be achieved, for instance, by inserting the attacker's SIM card into the victim's UE. However, if the exploit can be triggered before authentication, it would work with any SIM card inside the UE, significantly increasing the impact of the vulnerability. As it was described in Table~\ref{tab:results}, the vulnerabilities in all the COTS UEs happened before the authentication is established and therefore would work with any SIM card inside the UE.

Secondly, the cause of the crash is crucial. If the crash is due to a failed assert statement, the impact is limited to a Denial-of-Service (DoS) attack. On the other hand, if the crash results from a buffer overflow or use-after-free vulnerability, there is potential for escalation to remote code execution, which would have a much greater impact. Unfortunately, we do not know reasons of all the crashes to evaluate whether the certain crash is exploitable.

\subsubsection{Uplink fuzzing}
Assume that malicious packet \textit{X} triggers a vulnerability in the base station or core network. An attacker could exploit this vulnerability as follows:
\begin{enumerate}
    \item An attacker sets up a malicious UE.
    \item A malicious UE attempts to connect to the network and initiates a 4G/5G attach procedure.
    \item During the attach procedure, a malicious UE sends packet \textit{X}, thereby triggering the vulnerability.
\end{enumerate}
This uplink fuzzing scenario is straightforward and requires only one malicious UE to execute the attack.

All the arguments we made in the downlink scenario hold here as well. However, it is much easier for the attacker to pass the authentication stage successfully by buying an appropriate SIM card. Therefore, the vulnerabilities happening before and after the authentication can be considered equally severe. 
Regarding the cause of the crash, in case of the uplink fuzzing, we argue that even a DoS attack should be considered highly impactful as it could cause massive disruption of the entire network.

\section{Limitations}

\subsection{Fuzzing only the \textit{Attach procedure}}
In our framework, we rely on an open-source 4G/5G stack implementation to generate benign packets. However, these open-source projects cover only a limited subset of the messages defined in the 3GPP specifications, whereas commercial off-the-shelf (COTS) UEs are expected to implement all such messages. This limitation is one of reasons why our current work focuses exclusively on fuzzing the \textit{Attach procedure}. Triggering the \textit{Attach procedure} using open-source tools is relatively straightforward, while other procedures can be much more challenging to initiate. However, the techniques described in this paper can be adapted to fuzz other procedures as well. Despite this, we have argued that identifying vulnerabilities in the initial messages of the \textit{Attach procedure} (before authentication) is more impactful than finding vulnerabilities after authentication.

\subsection{No support for Apple devices}
Currently, our framework is unable to fuzz Apple devices due to two main constraints: first, Apple restricts access to the modem by not allowing the use of AT commands, and second, toggling Airplane Mode requires user interaction, as there is no iOS equivalent of ADB for automating such actions. To fuzz an Apple device, one could use a jailbroken device, which would allow direct communication with the modem. However, during our research, we did not have access to a jailbroken device and were therefore unable to test these techniques.

\section{Related work}
In recent years, a significant number of research papers have been published in the field of wireless protocol testing.
In this section, we first highlight studies that focus on over-the-air protocol testing, which closely aligns with our work and compare their approach with ours. Following that, we discuss other works that employ approaches which can be considered complementary to ours.

\subsection{Over-the-air wireless protocol testing}
Garbelini et al.~\cite{Greyhound_for_4G_5G} proposed a $4$G/$5$G fuzzing framework built on top of OpenAirInterface~\cite{OAI} (OAI) that supports fuzzing on multiple layers, similar to our research. 
Initially, we attempted to use OAI as the \textit{protocol stack implementation} but found it extremely unstable, as the majority of malicious packets caused crashes in the protocol stack. The need to recover from these crashes would slow down the fuzzing process and complicate the evaluation of the coverage-based algorithm. Consequently, we chose srsRAN 4G as our \textit{4G protocol stack implementation}.
Additionally, Garbelini et al. employed a different fuzzing method. They relied on user-defined test cases to guide fuzzing and on a dynamically built state machine to detect improper transitions. In contrast, our fuzzing framework does not require any human input to guide the fuzzing process, instead relying on an automatic coverage-driven algorithm.
The authors initially reported no vulnerabilities in COTS User Equipment (UEs), noting that ``more testing and experiments are required to identify new UE vulnerabilities."

However, in December 2023, they published a vulnerability disclosure report~\cite{5ghoul} detailing 12 vulnerabilities discovered during the 5G attach procedure on COTS devices. Among the tested devices we share one common target, namely Quectel RM500Q-GL modem, in which they identified one vulnerability (CVE-2023-33042). As the authors do not provide a firmware version of the modem -- only the firmware release date -- it is unclear whether our tests used the same version. Nevertheless, our testing uncovered five additional crashes (one in 4G and four in 5G) in the same modem.

The same team of authors published an extended version of their fuzzing framework~\cite{U-Fuzz}, which supports fuzzing various IoT protocols (CoAP, Zigbee, and 5G NR). 
However, the core idea behind their fuzzing algorithm remained similar to that in~\cite{Greyhound_for_4G_5G}.

In their other works, Garbelini et al. proposed fuzzing frameworks for WiFi~\cite{Greyhound} and Bluetooth~\cite{SweynTooth}, utilizing a different methodology to guide fuzzing, specifically the Particle Swarm Optimization (PSO) algorithm~\cite{PSO}. Their approach involved minimizing a cost function that included the number of transitions in a state machine, anomaly period, anomaly count, and iteration time, but not code coverage. For packet mutation, they assigned two probability values: one for the likelihood of mutating the packet and another for mutating each field within the packet. They chose not to assign probabilities to every field individually, as this would excessively expand the search space for the PSO algorithm.
In contrast, our framework employs a more fine-grained approach to packet mutation, where each field has its own probability of being mutated. Additionally, these probabilities are dynamically adjusted constantly seeking for new coverage.

Among other works aimed at cellular protocol fuzzing, Potnuru et al.~\cite{BERSEKER} focused on mutation- and generation-based fuzzing on the RRC layer and found two vulnerabilities in the open-source 4G implementations -- srsLTE (former name of srsRAN) and openLTE~\cite{OpenLTE}. However, their approach does not utilize feedback-driven fuzzing; instead, their mutation strategies are fixed and do not adapt based on the behavior of the \textit{DUT}.

Several other studies involve extensive manual effort to create test cases by analyzing the relevant 3GPP documents. Park et al.~\cite{DoLTEst} proposed a negative testing framework to uncover  non-standard-compliant behaviours in 4G implementations. Similarly, Kim et al.~\cite{TouchingUntouchable} generate test cases based on manual analysis of the 4G standard. Khandker et al.~\cite{ASTRA-5G} introduced a more automated approach to testing 5G SA devices, incorporating Large Language Models (LLM) in the evaluation stage. However, they found that LLMs were unable to perform well with the test cases that accumulate the parameters that are not mentioned in the specification or violate multiple security parameters at the same time. Additionally, they noticed that often LLM fabricates answers referring to non-existent parts of the specification.

A more successful application of LLMs in fuzzing was demonstrated by Wang et al. in their recent work~\cite{LLMIF}, where they effectively leveraged LLMs to fuzz Zigbee devices. First, they augmented the LLM with knowledge from the Zigbee specification and utilized it to structure information about the different message types, which are otherwise difficult to interpret as they are described in natural language. This structured information was then used to intelligently generate and mutate Zigbee packets. Finally, they monitored the device under test's (DUT) response codes and employed the LLM to perform conformance testing with the Zigbee specification. Their approach significantly outperformed all state-of-the-art fuzzers, demonstrating over a $50\%$ improvement in code coverage of an open-source Zigbee stack implementation. Furthermore, they uncovered 11 vulnerabilities in real-world devices. Notably, this was one of the few wireless protocol fuzzing studies that utilized code coverage metrics to evaluate their framework. We believe that analyzing coverage is a crucial step in enhancing fuzzers, and we strongly encourage future researchers to incorporate it into their work wherever possible.

\subsection{Other works}
The works of Maier et al.~\cite{BaseSAFE} and Fang et al.~\cite{EmulRL_4G} employ emulation-based fuzzing. Ruge et al.~\cite{Frankenstein} leverage advanced firmware emulation to fuzz Bluetooth devices and uncovered a chip-independent RCE in Android devices.
Emulation-based fuzzing is generally more efficient than over-the-air fuzzing because the fuzzing iterations are faster, and coverage information can be directly obtained from the DUT. However, emulating a device is a challenging task and does not scale well. 

Grossi et al.~\cite{Exploit_Modern_Smartphone},~\cite{Over_The_Air_Baseband_Exploit} and Komaromy,~\cite{CVE-2022-21744},~\cite{CVE-2023-21517} demonstrated how to achieve over-the-air remote code execution (RCE) on 4G and 5G smartphones. In these works, the authors utilized reverse-engineering techniques to identify vulnerabilities that were ultimately escalated to RCE. Fuzzing techniques can complement reverse engineering by automatically locating vulnerable parts of the code.

Additionally, Hussain et al.~\cite{5GReasoner, LTEInspector} proposed frameworks for the formal verification of $4$G and $5$G control-plane protocols targeting design flaws in the protocol itself. In contrast, our work focuses on identifying vulnerabilities at the implementation level.

\section{Conclusion and Future Work}
In this paper, we proposed a framework for fuzz-testing 4G and 5G systems during the \textit{Attach procedure}. Our fuzzer supports both uplink and downlink fuzzing for 4G, as well as downlink fuzzing for 5G. It is capable of fuzzing packet fields across all layers except the physical layer.

Compared to existing works, our framework introduces two novel concepts: (i) using code coverage to guide packet manipulation for wireless protocol fuzzing, and (ii) estimating code coverage when it cannot be directly extracted from the \textit{DUT}. We demonstrated that both approaches lead to better coverage of the \textit{DUT} than using an optimized \textit{Random fuzzer}.
During our fuzzing tests, we discovered vulnerabilities in ten out of twelve tested COTS UEs and in all of the srsRAN 4G instances.

Our framework is designed to be generic, meaning it can be adapted to fuzz other network protocols. It is also user-friendly, requiring no human intervention to prepare and run the fuzz tests. We leave further optimizations building upon our framework for future research, which we support by open-sourcing the code of our fuzzing framework.

For future work, integrating LLMs into fuzzing is a promising and emerging area with significant potential to extend this research. Another ambitious direction would involve combining over-the-air fuzzing (to identify vulnerabilities) with reverse engineering (for escalation of vulnerabilities to RCE). This would increase the impact of the found vulnerabilities and motivate manufacturers to enhance the security of their products.

\bibliographystyle{ieeetr}
\bibliography{references}

\end{document}